\documentclass[prl,aps,twocolumn,floats,nofootinbib,superscriptaddress]{revtex4-1}
\usepackage{graphicx}
\usepackage{dcolumn}
\usepackage{bm}
\usepackage[colorlinks=true,urlcolor=blue,citecolor=blue]{hyperref}
\usepackage{float}
\usepackage{subfig}

\usepackage{amsmath}
\usepackage{amssymb}
\usepackage{float}
\usepackage{physics}
\usepackage{comment}
\begin{document}


\title{Discrete-time crystal in periodically driven quantum Sherrington-Kirkpatrick model}

\author{Aarya Bothra}
 \affiliation{Department of Physical Sciences, Indian Institute of Science Education and Research Kolkata, Mohanpur 741252, India}%

\author{Arti Garg}%
 \affiliation{Theory Division, Saha Institute of Nuclear Physics, 1/AF Bidhannagar, Kolkata 700064, India}
\affiliation{Homi Bhabha National Institute, Training School Complex, Anushaktinagar, Mumbai 400094, India}

\begin{abstract}
  Discrete time crystals (DTC) have emerged as a significant phase of matter for out-of-equilibrium many-body systems. In this work, we study how random long-range interactions contribute to the stability of the DTC phase. Generally, a stable DTC phase is believed to be realized in disordered systems with short-range interactions. Here, we explore a periodically driven quantum Sherrington-Kirkpatrick (SK) model of Ising spin-glass in which all spins are coupled to each other randomly. We investigate the possibilities of the DTC phase in the SK model within three different driving protocols and found that the quantum SK model exhibits a robust DTC phase, although systems with uniform long-range interactions can exhibit only prethermal DTC phase. Further, we demonstrate that disorder in the local $XY$ term or transverse field is crucial for stabilizing a broad DTC phase, despite random couplings in the SK model. Our analysis shows that the stability of the DTC phase is determined by the non-ergodic nature of the Floquet eigenstates.
\end{abstract}

\maketitle

\section{I. Introduction}
\label{sec-1}
Periodically driven quantum systems have been shown to host novel non-equilibrium phases of matter that are impossible to achieve in equilibrium systems. One of the most remarkable examples of a non-equilibrium phase of matter is a discrete-time crystal (DTC) that spontaneously breaks discrete time-translation symmetry (dTTS), resulting in persistent oscillations in physical observables. Although breaking time-translation symmetry~\cite{Wilczek2012} is forbidden for systems in thermal equilibrium~\cite{Bruno2013,Nozieres2013,Watanabe2015}, a non-equilibrium setting can vouch for it, such as the periodically driven quantum systems. Generic periodically driven systems are expected to absorb energy indefinitely and reach an infinite-temperature state~\cite{Alessio2014}. Strongly disordered quantum systems in the many-body localised (MBL) phase~\cite{Basko2006,Gornyi2005,Altman2015,Alet2018,Nandkishore2015,Abanin2019,Sierant2025} can avoid heating when driven periodically~\cite{Abanin_Wojciech2016,Ponte_Chandran2015,Ponte_Papic2015}.
This enables achieving a DTC phase in periodically driven (Floquet) MBL systems~\cite{Khemani2016,Else2016,Ashvin2017,Choi2017,Khemani2021_comment,Ippoliti2021,BikunLi2020},  although clean systems can have at most a prethermal DTC 
phase~\cite{Kyprianidis2021,Beatrez2023,Stasiuk2023,Jiang2024,Yokota2025}. 

DTC phase has been experimentally realized in various systems such as trapped ions \cite{Zhang2017}, diamond nitrogen-vacancy centers \cite{Choi2017}, NMR experiments \cite{Rovny2018,Rovny2018_2,Pal2018} and more recently Rydberg atom arrays~\cite{Liu2024_higher}, superconducting qubits~\cite{Frey2022} and spin-based quantum simulators~\cite{Randall2021} which has opened new frontiers in quantum many-body physics.
In a Floquet system with a time-periodic Hamiltonian $H(t)=H(t+T)$, a DTC spontaneously breaks the dTTS of the drive such that physical observables show periodic oscillations with a period $mT$, that is an integer multiple of the driving period with $m>1$, corresponding to sharp subharmonic response in the frequency domain~\cite{Khemani2019,Ippoliti2021,Zaletel2023}. This emergent periodicity is robust against imperfections, making time crystals distinct from conventional dynamical oscillations and perfect for applications in quantum technology~\cite{Estarellas2020}, quantum sensors~\cite{Moon2024_Sensing,Yousefjani2025} and quantum metrology~\cite{Montenegro2023}.



In this work we explore the role of random long-range interactions in stabilizing the DTC phase. 
Systems with uniform long-range interactions are known to host prethermal DTC phase with strong initial state dependence~\cite{Machado2020,Kyprianidis2021,ChangLyu2020,Pizzi2021_higher}. General understanding is that the DTC phase in disordered systems can be realized only if the system has short-range interactions~\cite{Ippoliti2021} which is based on the belief that the MBL phase can not survive in the presence of long-range interactions. Recent work on MBL in the presence of long-range interactions has shown that for uniform long-range spin-preserving interactions MBL can survive as long as the spin-flip interactions are short-range~\cite{Burin2015,Yao2014,Gutman2016,Tikhonov2018,Nag2019,Roy2019,Jana2024}, though random long-range spin-preserving interactions may have a delocalization tendency~\cite{Prasad2021,Yao2014}.
A question of immense interest is whether a DTC phase can be realized in long-range interacting models with random interactions, and in this work we provide an affirmative answer to this question. 

We investigate the possibility of DTC in the quantum Sherrington-Kirkpatrick (SK) model~\cite{Sherrington1975,Goldschmidt1990,Lancaster1997,Takahashi2007,Young2017} which has spin-preserving all to all random interactions. To the best of our knowledge, DTC phase has not been explored in this model before. We used two different protocols to add quantum dynamics in the classical SK model, introducing a transverse field  and introducing a local $XY$ term. We first explore the possibility of the MBL phase in the periodically driven quantum SK model. In both of the above mentioned protocols, we found that as the strength of the random transverse field or $XY$ term is increased, a transition from non-ergodic Floquet MBL phase to an extended phase which has Poissonian statistics for eigen spacings takes place. We demonstrate that it is possible to stabilize a robust DTC phase in the driven quantum SK model in the parameter regime where the system is non-ergodic. 


We explore various driving protocols for periodically driven SK model with a two step drive. Starting from a random initial state, the first step does a spin flip for all the spins and the second step provides time evolution w.r.t the quantum SK model. If the quantum dynamics is induced by a transverse field, $h$, (uniform or random) then the quantum SK model shows a non-ergodic-DTC phase below a threshold value of the transverse field $h^\star$. 
On further increasing the field, the DTC order melts. 
The width of the DTC phase is much broader for the case of random transverse field  in comparison to that for the uniform transverse field. A qualitatively similar phase diagram is obtained in the driving protocol where in the second step of the drive, quantum dynamics in the SK  model is induced through the $XY$ coupling on the nearest-neighbour bonds instead of a transverse field. For random $XY$ couplings, $g_{ij} \in [0,g]$, the width of the DTC phase in the $g-J$ plane is broader compared to that for the case of uniform XY coupling. 

Our study shows that disorder in the spin- preserving long-range interactions is crucial for stabilizing a DTC phase. Along with it, disorder in the transverse field and $XY$ couplings also contribute to the stability and width of the DTC phase. Secondly, the nature of the quantum dynamics plays a significant role in the stability of the DTC phase; with interesting difference in long time dynamics of the system with random $XY$ term in comparison to the random transverse field. Interestingly, small random $XY$ terms can induce a DTC phase even in systems with uniform power-law interactions.

The rest of the manuscript is organised as follows. In section II, we discuss the delocalization to MBL transition in periodically driven quantum SK model for two driving protocols. In section III, we present results for time evolution dynamics for various driving protocols and provide evidence in support of the DTC phase in three different driving protocols and we conclude our discussions in  section IV.  

\section{Model and Driving Protocols}
\label{sec-2}
We study periodically driven quantum SK model in which all Ising spins are coupled to each
other, with couplings themselves being random. The first driving protocol that we consider is described by the time- dependent Hamiltonian $H_f(t)$ defined on the interval $[0,T)$ of period $T=2$ such that $H(t+T)=H(t)$. The stroboscopic Floquet Hamiltonian has the form:\begin{equation}
H_f(t) = 
\begin{cases} 
H_1 = \frac{\pi}{2}\sum\limits_i \sigma_i^x, & 0 < t < 1 \\ 
H_2 = \sum\limits_{i<j}J_{ij}\sigma_i^z\sigma_j^z + \sum\limits_ih_i^x\sigma_i^x, & 1 < t < 2
\end{cases}
\label{d1}
\end{equation}
Here, $\sigma_i^z$ and $\sigma_i^x$ are the z and x components of Pauli spin matrices respectively. 
$H_2$ represents quantum SK model with the coupling term $J_{ij}$ distributed following a Gaussian distribution with mean and variance as $0$ and $J/\sqrt{L}$ respectively~\cite{Sherrington1975}. We work with open boundary conditions. $h_i^x$ is the local transverse field which induces quantum dynamics in the classical SK model of Ising spins. We consider the case of uniform field $h_i^x=h$ for all sites as well as the case of random transverse field $h_i^x \in [0,h]$.
In the second driving protocol, we introduce a $XY$ coupling among spins rather than a transverse field term in $H_2$ 
such that the floquet hamiltonian is given as:
\begin{equation}
H_f(t) = 
\begin{cases} 
H_1 = \frac{\pi}{2}\sum\limits_i \sigma_i^x,\\ \hspace{5.2cm} 0 < t < 1 \\ 
H_2 = \sum\limits_{i<j}J_{ij}\sigma_i^z\sigma_j^z + \sum\limits_ig_i(\sigma_i^x\sigma_{i+1}^x+\sigma_i^y\sigma_{i+1}^y),\\ \hspace{5.2cm} 1 < t < 2
\end{cases}
\label{d2}
\end{equation}
In this protocol, the spins are arranged in a 1D geometry (which defines the nearest-neighbor interactions), but also possess an infinite-range, random Ising coupling (the SK term). Such Hamiltonian structures—where local 1D interactions compete with global all-to-all couplings—are highly relevant to current experiments like cavity QED or trapped ions.
Once again, we shall study the cases of both uniform coupling $g_i=g$ for all sites and random couplings $g_i$ drawn from a uniform distribution $g_i \in [0,g]$.  Since the $XY$ term flips a pair of spins which are antiparallel, it is not effective on a fully polarised initial state but for any random initial state, one can analyse unitary time evolution with $H_f(t)$. Our numerical analysis shows that the DTC phase obtained in this case is broader in $J-g$ plane compared to that obtained in the first driving protocol where a transverse field is used to introduce quantum dynamics. 
In both these protocols, we associate the robustness of the observed DTC phase with the properties of eigenstates of $H_f$. 


The third driving protocol that we studied is described below:
\begin{equation}
H_f(t) = 
\begin{cases} 
H_1 = (\frac{\pi}{2}-\epsilon)\sum\limits_i \sigma_i^x, & 0 < t < 1 \\ 
H_2 = \sum\limits_{i<j}J_{ij}\sigma_i^z\sigma_j^z, & 1 < t < 2
\end{cases}
\label{d3}
\end{equation}
Here for $\epsilon=0$ period doubling oscillations in physical observables are obvious, but we show that the DTC phase survives for a finite range of $\epsilon$ in the SK model, while it is absent in the system with uniform power-law interactions. 

\section{ II. Many-body localization in periodically driven Sherrington-Kirkpatrick model}
In this section, we explore Floquet-MBL in periodically driven quantum SK model.
The Floquet unitary operator for the driving protocol in Eqn.~(\ref{d1},\ref{d2}) is given by:
\begin{equation}
U_F=exp(-iH_2)exp(-iH_1)
\label{U}
\end{equation}
It corresponds to periodically driving the system with period $T=2$ and frequency $\Omega=2\pi/T$. In the following, we use the exact diagonalization of the Floquet unitary in Eq.~\ref{U} and obtain the eigenspectra $U_F|\Psi_n\rangle=e^{-iE_n T}|\Psi_n\rangle$ where $E_n$ are the eigenphases and $|n\rangle$ are the Floquet eigenstates. The eigenphases are restricted to the first Floquet zone $-\Omega/2 \le E_n \le \Omega/2$. 
Below we explore the Floquet MBL and delocalization transition first  in the driving protocol-1 followed by the analysis in the driving protocol-2. 

\begin{figure*}[t]
{\includegraphics[width=2.0\columnwidth]{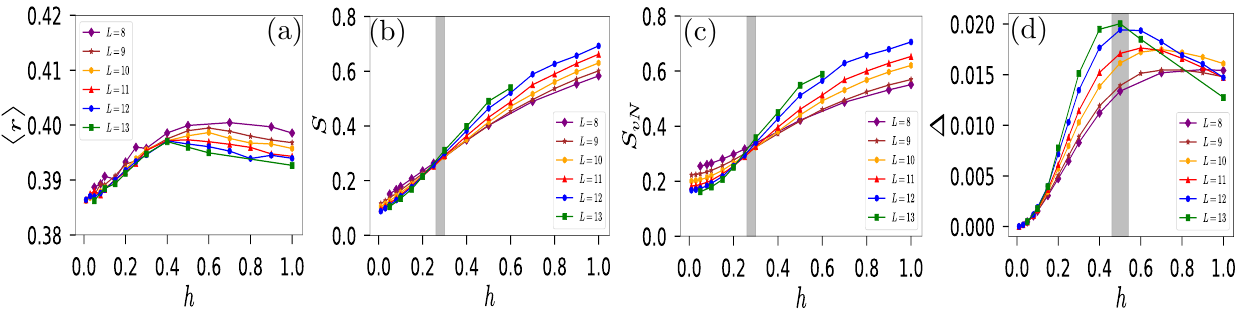}}

    \caption{MBL analysis of the periodically driven SK model under driving protocol-I with $J=2.0$ and random $h_i\in [0,h]$. System-size dependence of the disorder averaged (a) level spacing ratio $\langle r \rangle$, (b) Shannon entropy normalized by $ln(2^L)$, $S$, (c) bi-partite entanglement entropy normalized by $ln(2^{L/2})$, $S_{vN}$ and (d) variance of the bi-partite entanglement entropy normalized by $[ln(2^{L/2})]^2$, $(\Delta)$ calculated from the full quasienergy shown as a function of $h$. Disorder averaging was performed over $2000 (L=8)$, $3000 (L=9)$, $3000 (L=10)$, $1100 (L=11)$, $400 (L=12)$ and $100 (L=13)$ independent disorder configurations. The shaded grey region denotes the MBL transition regime inferred from the finite-size analysis.}
    \label{MBL_F1}
\end{figure*}
{\subsection{Many-body localization in driving protocol-I}}
We investigate several physical quantities to explore the Floquet MBL and delocalization transition first in the driving protocol-1. Level spacing ratio(LSR) is a conventional diagnostic to distinguish between ergodic and non-ergodic phases. Level spacing ratio is defined as $r_n=\frac{min(\delta_n,\delta_{n+1})}{max(\delta_n,\delta_{n+1})}$ with $\delta_n=E_{n+1}-E_n$ at a given eigenenergy $E_n$ of the quantum Hamiltonian. The distribution of energy level spacings is expected to follow Poisson statistics in the non-ergodic phase with disorder-averaged $\langle r \rangle \approx 0.386$ while it follows circular orthogonal ensemble for the ergodic phase with $\langle r \rangle \approx 0.529$~\cite{Alessio2014,KhemaniFlMBL}. In panel[a] of Fig.~\ref{MBL_F1}, we have shown LSR averaged over the entire spectrum as a function of $h$ in the regime $h <J$. For the entire regime of $h$, the LSR remains close to the Poisson value. Although it increases slightly as $h$ increases, there is also a significant decrease with increase in system size keeping the average LSR always close to the Poissonian value.   Thus, in the regime of interest $h < J$, the system is non-ergodic. 

We also calculated the Shannon entropy for every eigenstate of $U_F$. Shannon entropy, which provides a clean way of analysing the localization properties of the eigenstates of a system, is defined as $S(E_n)=-\sum_{i=1}^{2^L}|\psi_n(i)|^2 \log|\psi_n(i)|^2$. For a many-body state that gets contributions from all the basis states in the Fock space (and is normalized), $S(E_n)\sim \log(2^L)$. Normalized $S$ is equal to one for a maximally delocalized state, while for a localized state that gets significant contribution only from a limited set of the basis states, $S$ vanishes to zero in the thermodynamic limit. Fig.~\ref{MBL_F1} shows the average normalized Shannon entropy $S$ obtained by averaging over all the eigenstates in the spectrum as well as over a large number of independent disorder configurations. For a fixed strength $J$, as the strength of the random transverse field, $h$, increases, the normalized Shannon entropy increases from zero to one, showing the delocalization induced by the transverse field. The MBL transition regime has been represented by a gray bar in Fig.~\ref{MBL_F1}. For $h < h_c$, $S$ decreases as the size of the system increases, while it increases with $L$ approaching one for $h > h_c$.

Next we consider the bipartite entanglement of the half-chain using the von Neumann entanglement entropy $S_{vN}$. We divide the system into two equal sublattices $A$ and $B$ and calculate $S_{vN}$ for each eigenstate of the Floquet operator $U_F$ with $S_{vN}(E_n) = -Tr[\rho_n^A \log\rho_n^A]$ and $\rho_n^A = Tr_B[|\Psi_n\rangle \langle \Psi_n|]$. For small values of the transverse field, $S_{vN}$ obeys area-law  while for large values of $h$ each half chain thermalizes to infinite temperature phase~\cite{Alessio2014} in which $S_{vN}$ shows the volume-law behaviour for all the eigenstates of $U_F$ approaching the Page value~\cite{Page} for large enough system size. Panel[c] in Fig.~\ref{MBL_F1} shows normalized entropy $S_{vN}/\ln(2^{L/2})$ averaged over the entire Floquet spectrum as well as over a large number of independent disorder configurations. In the MBL phase with $h < h_c$, normalized $S_{vN}$ decreases as $L$ increases while in the delocalized phase $S_{vN}$ increases with L approaching one in the infinite size limit. 
We also analysed the variance of the distribution of entanglement entropy $\Delta=\langle S_{vN}(E_n)^2 \rangle-\langle S_{vN}(E_n)\rangle^2$ where $\langle \rangle$ represents averaging over entire spectrum as well as a large number of independent disorder configurations. $\Delta$ shows a peak as a function of $h$ but the peak position $h_{c2}(L) > h_c(L)$ though it shifts towards $h_c$ as $L$ increases.

Though Shannon entropy and entanglement entropy show a clear transition from Floquet MBL to a delocalzed phase at $h_c$, the LSR remains close to the Poissonian value even for $h > h_c$ indicating the non-ergodic nature of the system even for $h > h_c$. Though we can not rule out the possibility of this being an artifact of the folding of states in the Floquet spectra. 

\begin{figure*}[t]
    \centering
    {\includegraphics[width=2.0\columnwidth, height=0.8\columnwidth]{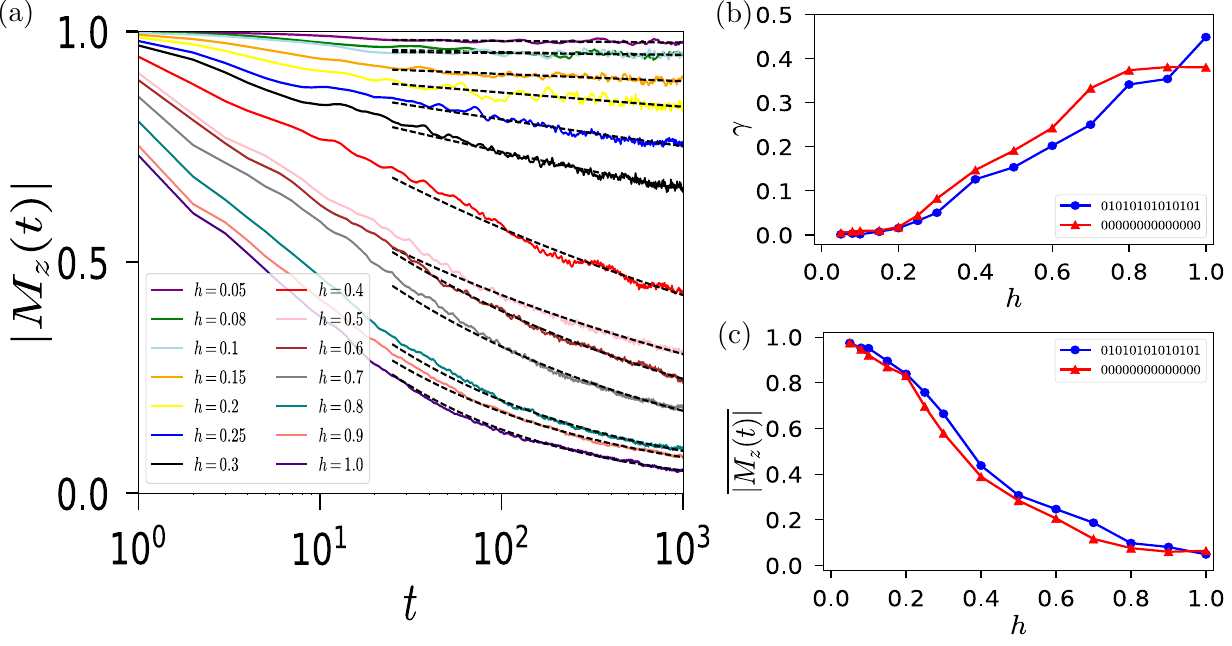}}

    \caption{(a) Absolute value of the average magnetisation $|M_z(t)|$ for the Neel initial state $|01010101010101\rangle$ $(L=14)$ in the periodically driven SK model under driving protocol-I with $J=2.0$ and random $h_i \in [0,h]$. Data shown was averaged over $50$ independent disorder configurations. (b) Decay exponent, $\gamma$ obtained from a linear fit to $|M_z(t)|$ over the time interval $t=25-1000$ on a log-log scale for two different initial states - the fully polarised and the Neel state. (c) Time-averaged absolute magnetisation, $\overline{|M_z(t)|}$ evaluated over the interval $t=900-1000$ for the same initial states. The two initial states exhibit similar magnetisation decay behaviour, indicating that the relaxation dynamics are largely independent of the choice of initial state.}
    \label{M_h1}
\end{figure*}

{\bf Quantum Quench Dynamics:}
To further explore the non-ergodicity in the periodically driven SK model, we also computed spatially averaged dynamical quantities for certain initial states. For an initial state $|\Psi_0\rangle$, we calculated $|M_z(t=nT)| = |\frac{1}{L}\sum_{i=1}^{L} \langle \Phi_0|\sigma_i^z(t)\sigma_i^z(0)|\Phi_0\rangle$. 
The left panel of Fig.~\ref{M_h1} shows $|M_z(t)|$ for a fully polarised initial state. For small values of $h$, magnetization does not decay with time which shows that system is non-ergodic and has long-term memory of the initial state. As $h$ increases, the decay rate increases which shows the loss of memory and transition into the ergodic phase. In the long time limit, magnetization shows a power-law decay $|M_{z}(t)| \sim t^{-\gamma}$ with $\gamma \sim 0$ (within error-bars) for $h < h_c$ while for $h \ge h_c$, $\gamma$ increases monotonically with $h$ approaching the diffusive limit for $h \sim 1.0$ for $J=2.0$. Panel [b] in Fig.[\ref{M_h1}] shows $\gamma$ vs $h$ for the fully polarised initial state as well as the Neel state. For both states $\gamma$ has a very similar trend. Panel (c) shows the value of $|M_z|$ in the long time limit. 

{\subsection{Many-body localization in driving protocol-II}}
In the second driving protocol, we introduce $XY$ coupling among nearest-neighbour spins rather than a transverse field term in $H_2$ as shown in Eqn.~\ref{d2}. 
\begin{figure*}[]
    {\includegraphics[width=2.0\columnwidth]{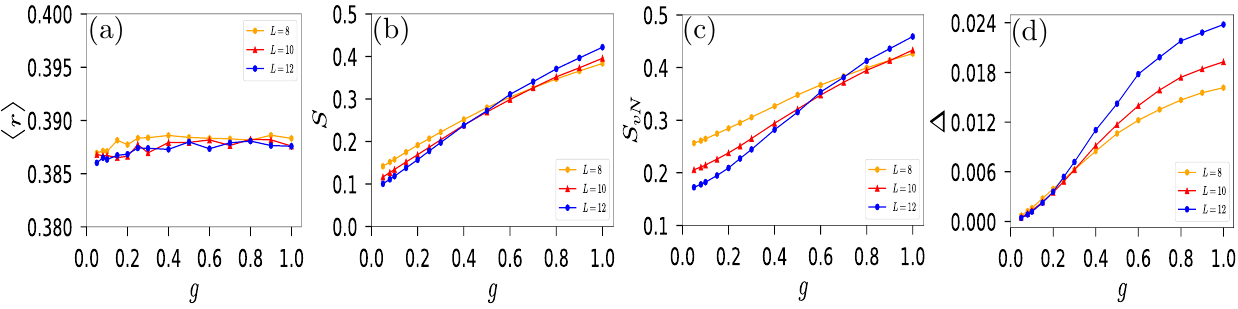}}

    \caption{MBL analysis of the periodically driven SK model under driving protocol-II with $J=2.0$ and random $g_i\in [0,g]$. System-size dependence of the disorder averaged (a) level spacing ratio $\langle r \rangle$, (b) Shannon entropy normalized by $ln(2^L)$, $S$, (c) bi-partite entanglement entropy normalized by $ln(2^{L/2})$, $S_{vN}$ and (d) variance of the bi-partite entanglement entropy normalized by $[ln(2^{L/2})]^2$, $(\Delta)$ calculated from the full quasienergy shown as a function of $h$. Disorder averaging was performed over $5000 (L=8)$, $3000 (L=10)$ and $800 (L=12)$ independent disorder configurations.}
    \label{MBL_F2}
\end{figure*}
For $J=0$, the random XY model on the nearest-neighbour sites, which can also be mapped onto the random hopping model for fermions, has a long interesting history~\cite{Lieb1961,Theodorou1976}. In the absence of SK interactions in $H_2$, the random $XY$ coupling is enough to give rise to localized states with anomalous behavior near the center of the band~\cite{Eggarter1978}. For any well-behaved distribution of spin-flip couplings, the density of states diverges as 
$\rho(E) \sim 1/|E\ln^3 |E||$ ~\cite{Dhar1980}, and the localization length diverges as $\xi(E) \sim |\ln|E||$. The state at zero energy is not conventionally extended but decays as $\Psi (r) \sim \exp(-\sqrt{r/r_0})$~\cite{Fleishman1977}. The effect of adding these random spin-flip terms to the all to all connecting random Ising term is interesting to explore in the context of Floquet-MBL. 
\begin{figure*}[t]
    {\includegraphics[width=2.0\columnwidth,height=0.8\columnwidth]{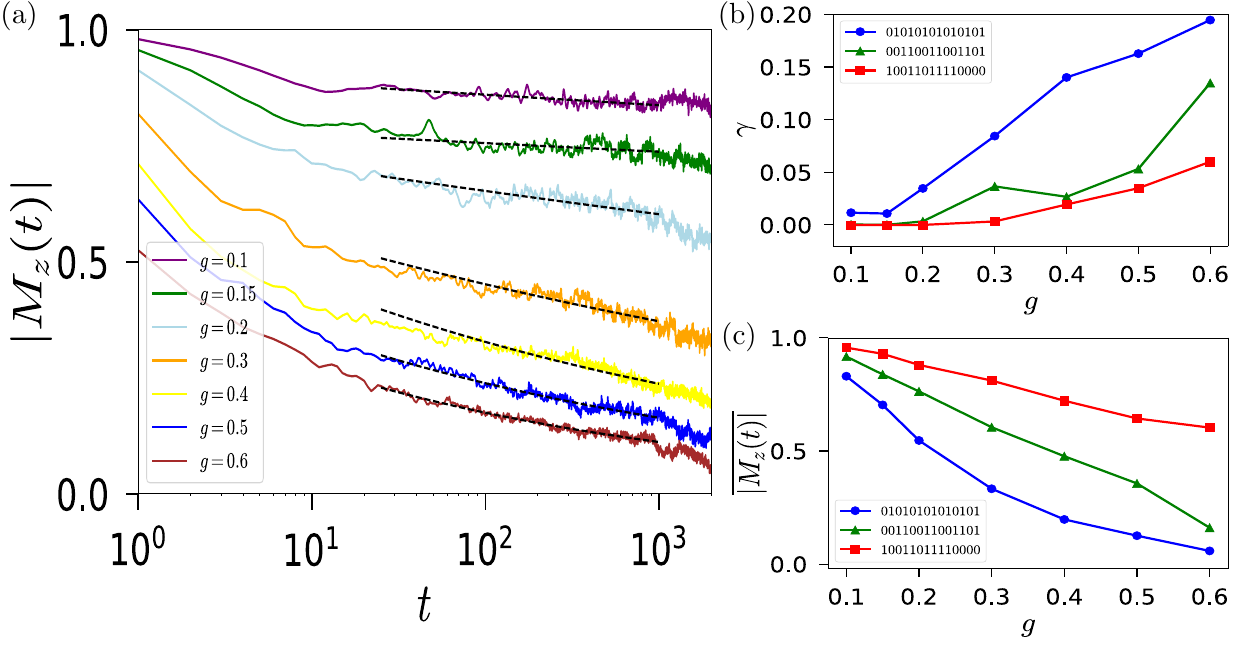}}

    \caption{(a) Absolute value of the average magnetisation $|M_z(t)|$ for the Neel initial state $|01010101010101\rangle$ $(L=14)$ in the periodically driven SK model under driving protocol-II with $J=2.0$ and random $g_i \in [0,g]$. Data shown was averaged over $50$ independent disorder configurations. (b) Decay exponent, $\gamma$ obtained from a linear fit to $|M_z(t)|$ over the time interval $t=25-1000$ on a log-log scale for three different initial states. (c) Time-averaged absolute magnetisation, $\overline{|M_z(t)|}$ evaluated over the interval $t=1900-2000$ for the same initial states. The initial states exhibit different magnetisation decay behaviour, with the decay rate increasing with the number of domain walls.}
    \label{Mz_g}
\end{figure*}

As shown in Fig.~[\ref{MBL_F2}] LSR of the Floquet spectrum corresponding to the system in Eqn.~\ref{d2} obeys Poissonian statistics for $g \le 1$ while the average of Shannon entropy as well as the bipartite entanglement entropy indicates a MBL to delocalization transition around $g_c\sim 0.6$. The variance of the bipartite entanglement entropy is peaked at a $g$ value larger than $g_c$. Thus, based on the analysis of various statistical quantities, it seems that the system remains many-body localized for a large range of $g$, at least up to $g \sim 0.6$ although LSR shows Poissonian statistics even for larger values of $g$. This implies that for $g > 0.7$  also the Floquet system is non-ergodic, but again the folding of eigenstates spoiling the LSR cannot be ruled out. 

Fig. ~[\ref{Mz_g}] shows the results of quantum quench starting from a Neel state. As $g$ increases, the system loses memory of the initial state, resulting in a faster decay of $|M_z(t)|$. The decay exponent $\gamma$ for three different initial states is shown in Panel [b]. For $g \le 0.15$, $\gamma \rightarrow 0$ for all the initial states indicating that the system is non-ergodic and localized having a strong memory of the initial state. As $g$ increases, $\gamma$ increases for all initial states, though staying in the sub-diffusive regime, but a strong initial state dependence is seen which was not there in the system in driving protocol-1. A similar trend is seen in the long time value of $|M_z(t)|$ shown in Panel[c] where for $g > 0.1$, a significant dependence on initial states is seen. 
\section{III. Discrete Time Crystal in Periodically Driven Sherrington-Kirkpatrick model}

\begin{figure*}[t]
    \centering
    (a)\subfloat{\includegraphics[width=0.63\columnwidth,height=0.5\columnwidth]{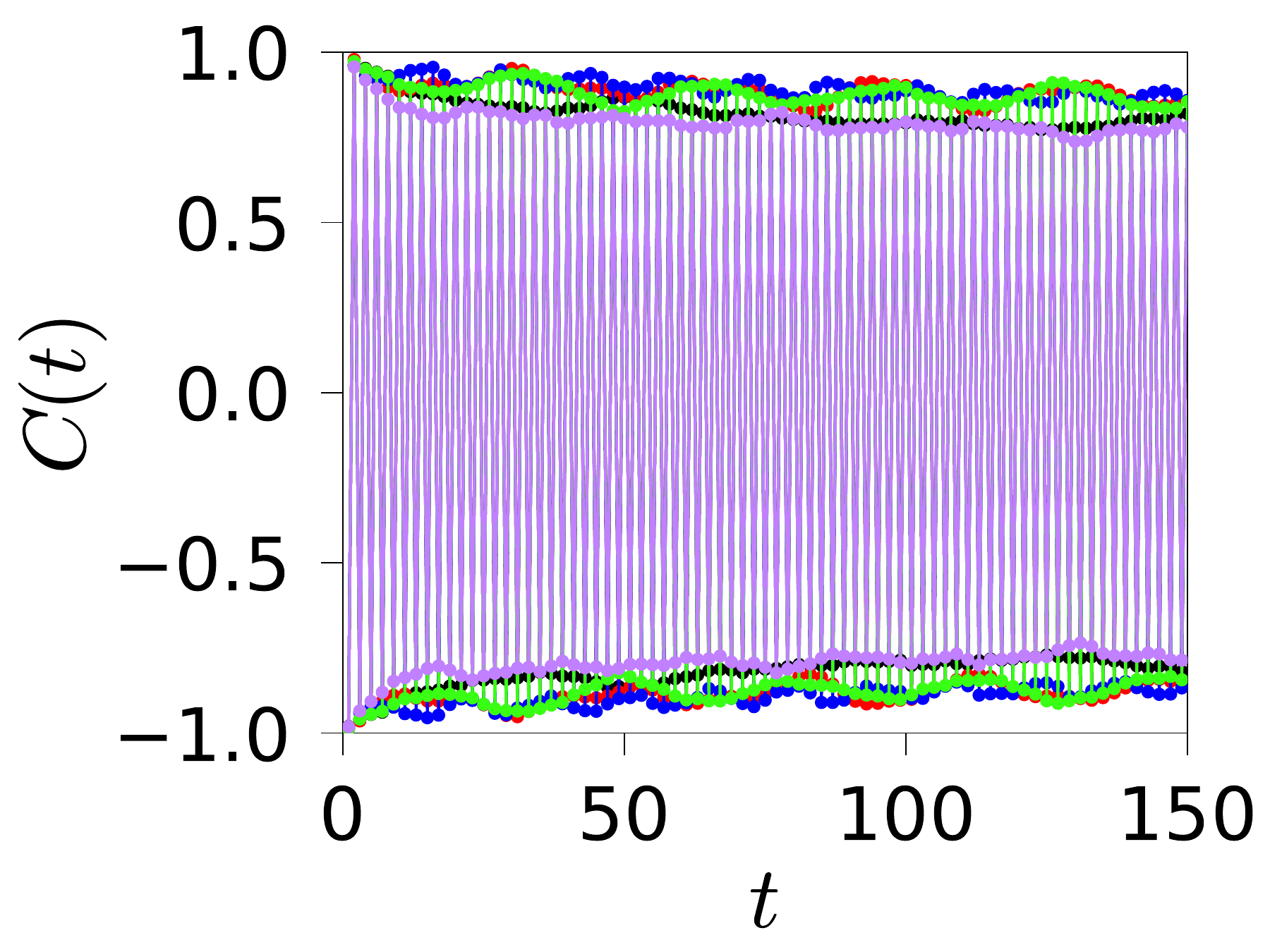}}
    (b)\subfloat{\includegraphics[width=0.63\columnwidth,height=0.5\columnwidth]{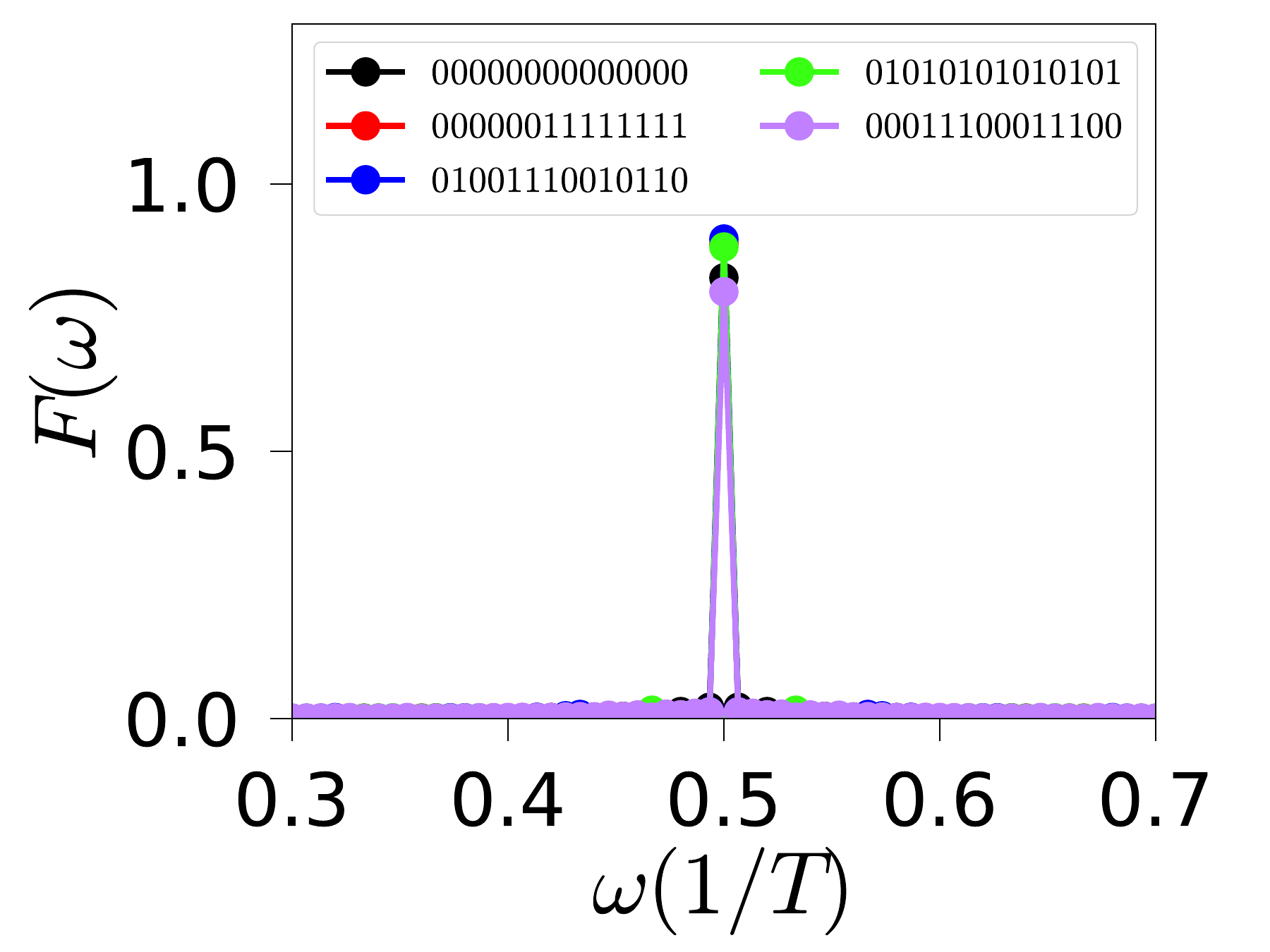}}
    (c)\subfloat{\includegraphics[width=0.63\columnwidth,height=0.5\columnwidth]{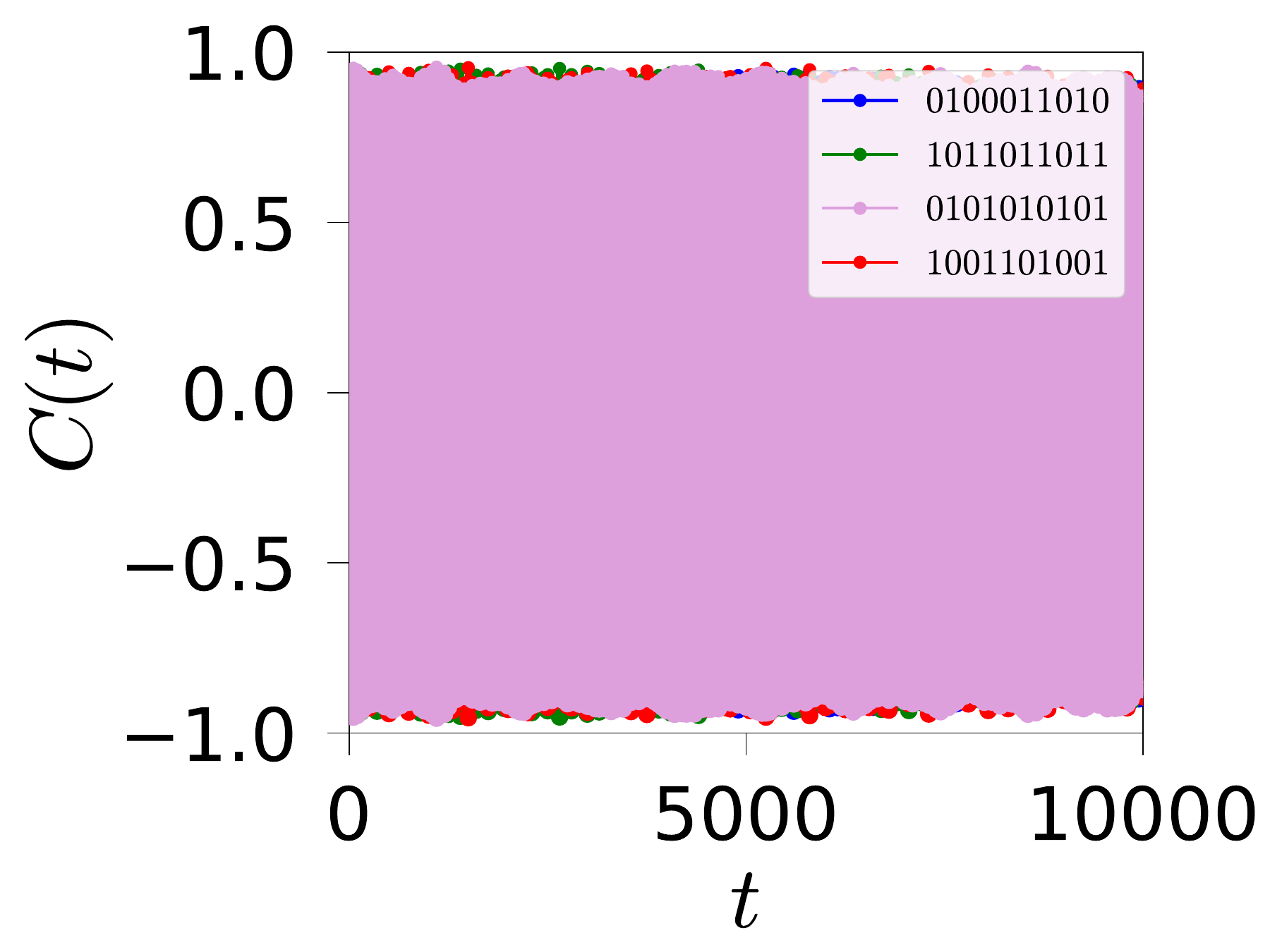}}

    \caption{(a)Autocorrelation function $C(t)$ for the periodically driven quantum SK model under driving protocol-I. For a large number of initial states studied, persistent oscillations of period $2T$ are observed in $C(t)$. (b) The Fourier transform of the correlation function, $F(\omega)$ clearly shows a singular peak at $\omega=0.5$. The data shown is for $J=2.0$ and random transverse field $h_i \in [0,h]$ with $h=0.2$ for a spin-chain of size $L=14$. For each initial state, averaging over 25 different disorder configurations was performed. (c) Long time $C(t)$ shown for different initial states of size $L=10$ at the same parameter value. Even at time steps up to $10^3$, minimal decay in $C(t)$ is seen highlighting the stability of the DTC phase.}
    \label{ct_h}
\end{figure*}

In this Section we explore the possibility of a DTC phase in periodically driven SK model. Starting from a random initial state $|\Psi_0\rangle$, we shall employ unitary time-evolution $U_f |\Psi_0\rangle= e^{-iH_2}e^{-iH_1}|\Psi_0\rangle$ on different initial states and calculate the disorder averaged spin-spin autocorrelation function $C_i(t)=\langle\psi_0|\sigma_{i}^z(t) \sigma_{i}^z(0)|\psi_0\rangle$ at stroboscopic time steps $t=nT$. An infinitely long-lived period-doubled oscillations in such autocorrelators of local operators characterize the DTC phase. Starting with a product state in the $\sigma_z$ basis, action of unitary $U_1= exp(-iH_1) = \prod\limits_{i=1}^L e^{-i(\frac{\pi}{2})\sigma_i^x}$ is that it induces a $180^\circ$ rotation and flips all spins where $U_1=\prod\limits_{i=1}^L -i\sigma_i^x$. This is followed by the time evolution under the Hamiltonian $H_2$. 

For $h=0$, since the eigenstates of $\sigma_z$ are also eigenstates of $H_2$, $U_2 = exp(-iH_2) = exp(-iE\{\sigma_i\})$ only provides a phase shift without any flipping of the spins. Thus, for $h=0$, the spins are flipped back when $U_1$ acts second time resulting trivially in period doubling dynamics. But the interesting question is to look for stability of this DTC phase of period $T=2$ for any finite non-zero $h$.  We demonstrate here from the direct time evolution of various random initial states that for the SK model, which has long-range interaction among spins as well as sufficient randomness in the spin-spin couplings to induce non-ergodicity in the system, it is possible to have a robust DTC phase for a finite range of $h$ for both the uniform and random transverse fields though the width of DTC phase is much broader in the case of random transverse field. Below we first present results for the driving protocol-I.

\begin{figure*}
    \centering
  \includegraphics[width=1.8\columnwidth]{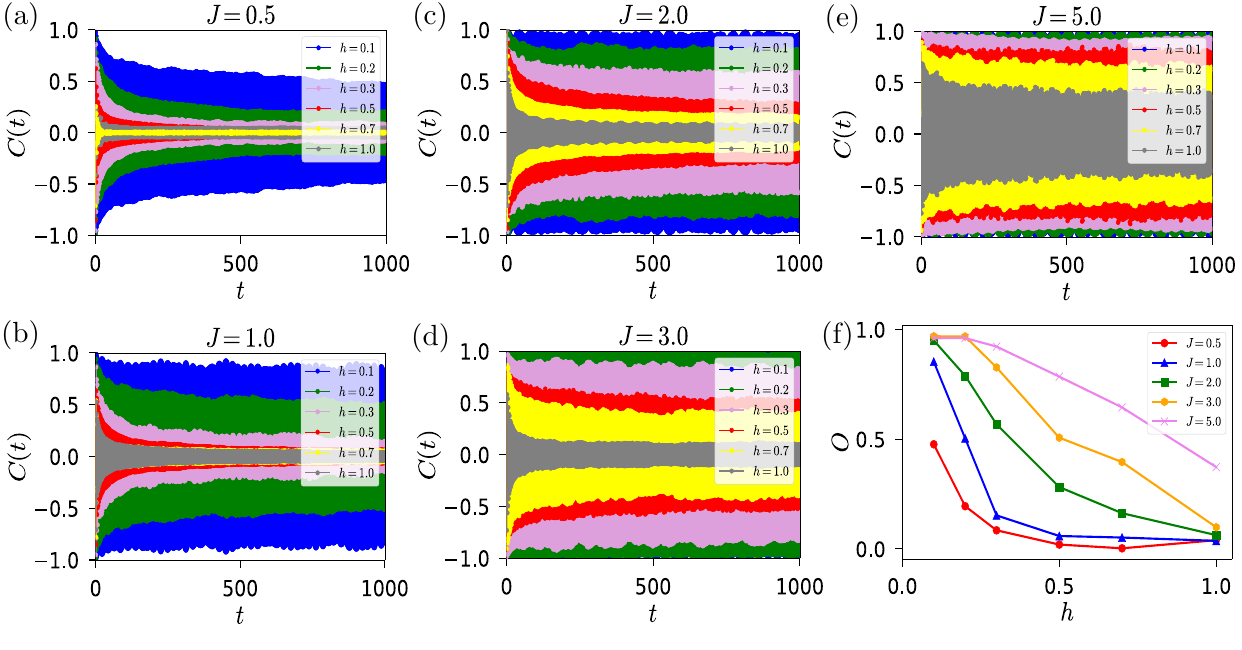}
  
    \caption{(a)-(e) Autocorrelation function $C(t)$ for the periodically driven quantum SK model under driving protocol-I for a range of $J$ values and random $h_i \in [0,h]$.  (f) Order parameter $O$ vs $h$ for various values of $J$. Correlation function has been calculated for one domain wall initial state $|00000011111111\rangle$,  $(L=14)$ and averaging was performed over $50$ independent disorder configurations.}
    \label{diffJ_ct}
\end{figure*}

\begin{figure*}
    \subfloat{\includegraphics[width=0.6\columnwidth,height=0.5\columnwidth]{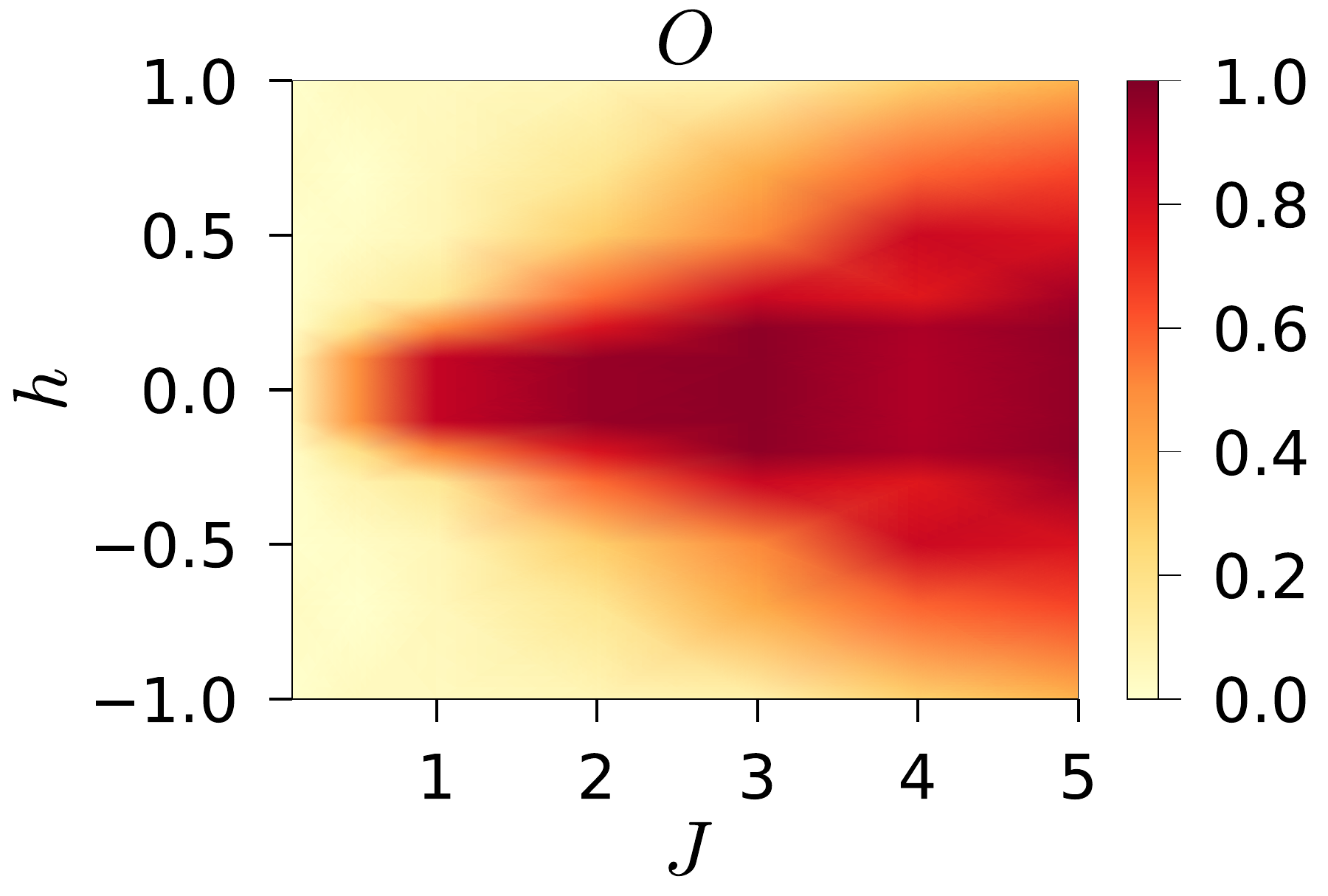}} 
     \includegraphics[width=0.53\columnwidth,height=0.5\columnwidth]{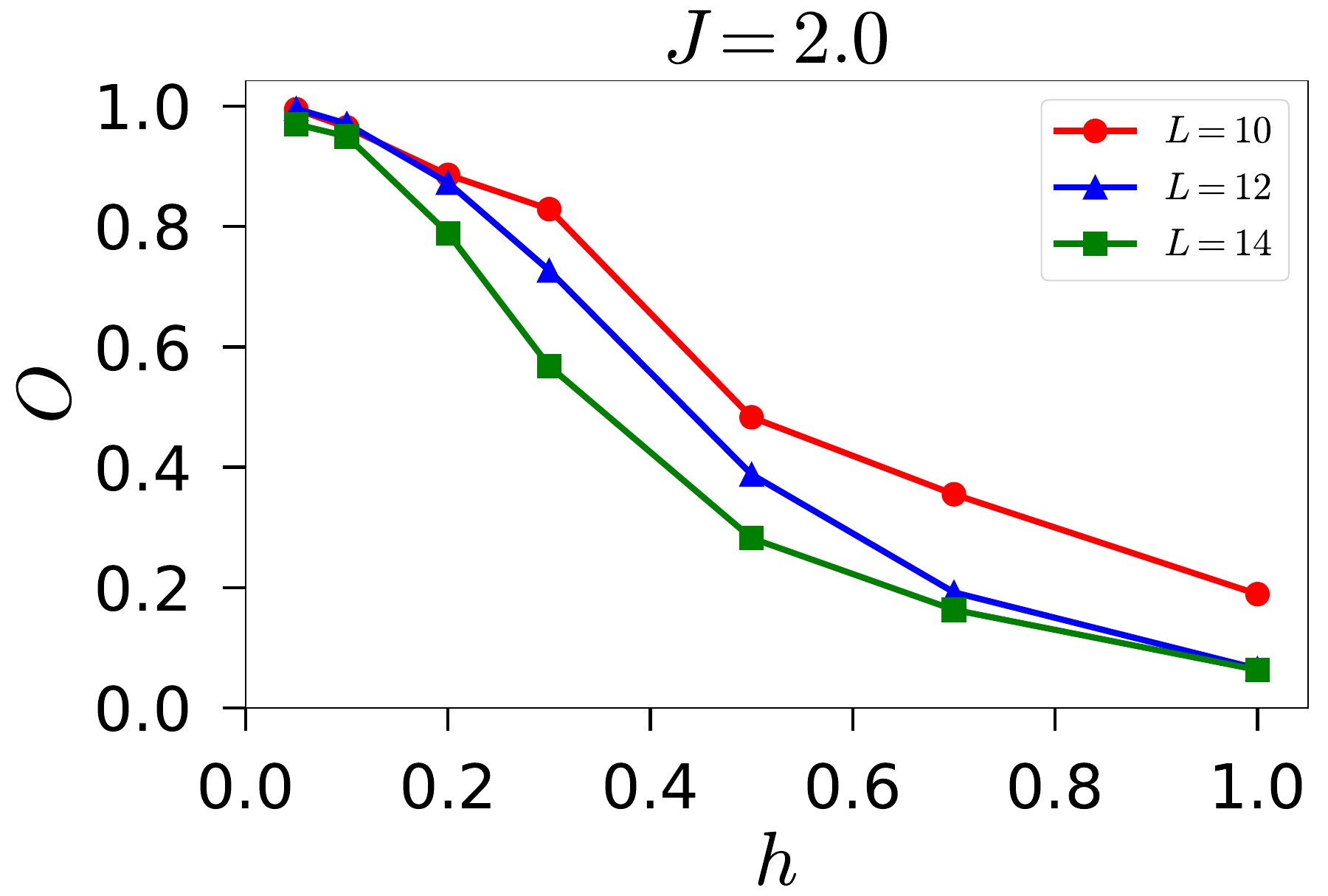}
     \includegraphics[width=0.53\columnwidth,height=0.47\columnwidth]{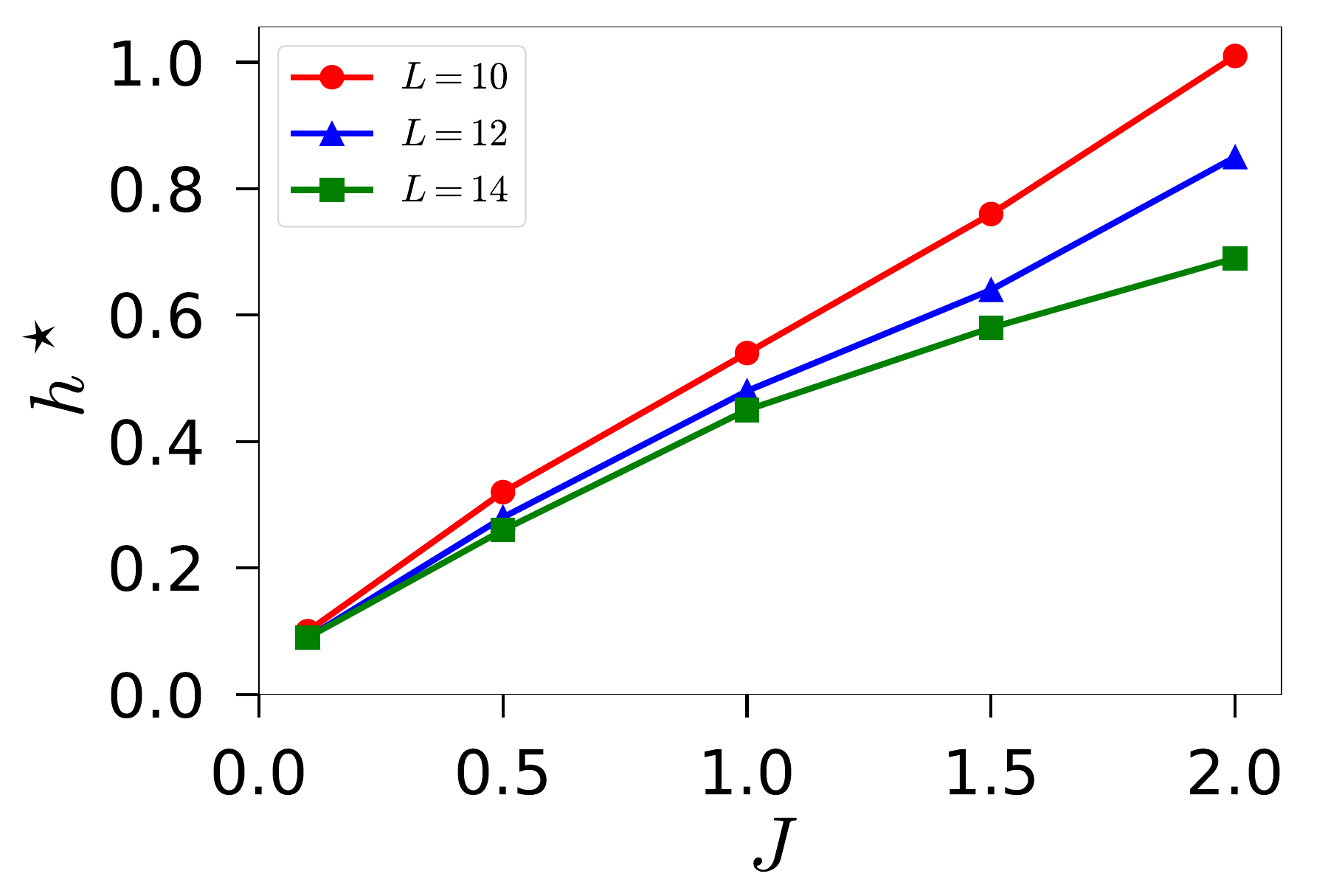}

    \caption{Order Parameter, $O$, calculated for the initial state $|00000011111111\rangle $ in the $J-h$ plane for periodically driven quantum SK model under driving protocol-I for $N=1000$. 
    Middle panel shows $O$ vs $h$ for various system sizes for $J=2.0$ while the last panel shows the critical field $h^\star$ vs $J$ for three systems sizes.
   The data shown has been averaged over $1000-50$ different disorder configurations for $L=10-14$.}
    \label{OP}
\end{figure*}

\subsection{Driving Protocol-I}
{\bf Auto-correlation Function and the DTC order-parameter}: Fig~\ref{ct_h}, shows the disorder averaged autocorrelation function $C(t)=\langle \Psi_0|\sigma_i^z(t)\sigma_i^z(0)|\Psi_0\rangle$ vs $t=nT$ calculated at $i=L/2$ for various initial states for the periodically driven quantum SK model under driving protocol-I.
Here we have shown $C(t)$ for $J=2.0$ and a weaker random transverse field $h=0.2$ where the Floquet system is non-ergodic. For all the initial states studied, $C(t)$ shows persistent periodic oscillations of period $2T$ for a long time. The data shown are for $t_{max}=150T$ but we have checked by going to larger times as well (shown in panel 3 of Fig.~\ref{ct_h}) that there is no decay of periodic oscillations for any initial state. 
We further calculate the Fourier transform $F(\omega)$ of $C(t)$ in the frequency space, which has a very sharp peak at $\omega=0.5W$ where $W=2\pi/T$ is the drive frequency. This shows that the periodically driven quantum SK model has the DTC phase of period $2T$ which breaks the discrete time translational symmetry of $H_f(t)$ in Eqn~\ref{d1}. 
\begin{figure*}[t]  
 \centering
\subfloat{\includegraphics[width=1.7\columnwidth]{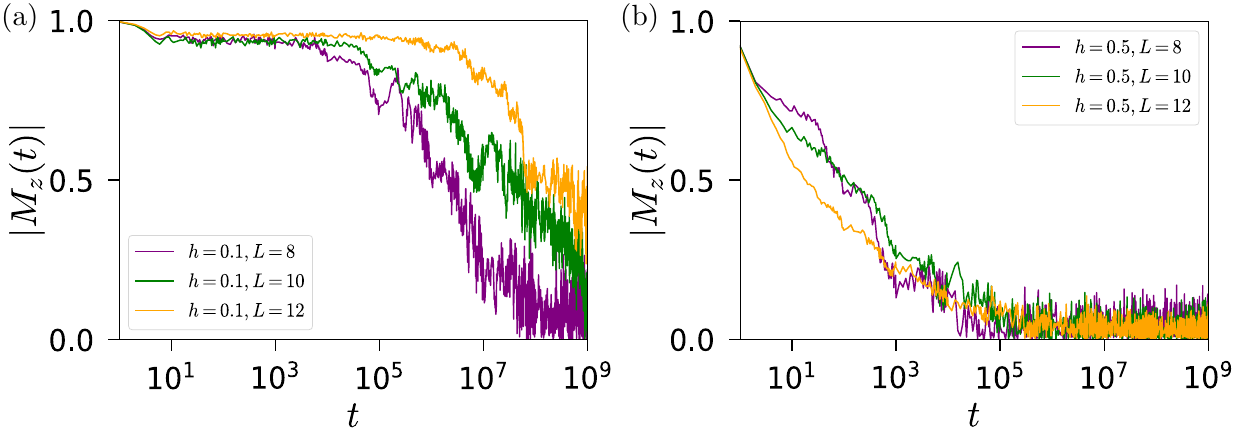}}

    \caption{Staggered magnetization ($|M_z(t)|$) vs $t$ for the Neel state as initial state for the driving protocol-1 with random transverse field $h$ for three systems sizes. The left panel shows the results for $h=0.1$ and the right panel shows the results for $h=0.5$. The data shown is for $J=2.0$ and has been averaged over 50 independent disorder configurations.}
    \label{M_longt_h}
\end{figure*}

We also explored the $J$ dependence of the DTC phase. Fig.~\ref{diffJ_ct} shows $C(t)$ for various values of $J$ for one domain wall initial state. Since the SK interaction term induces disorder and non-ergodicity in the system while the transverse field provide dynamics to delocalize the system, for smaller values of the SK interaction $J$, the correlation function decays at smaller values of $h$. 
In order to analyse the width of the DTC phase in the $h-J$ plane, we computed the order parameter $O$ for a random initial state, which is defined as follows,
\begin{equation}
\begin{aligned}
    O=\frac{1}{N}\sum_{n=1}^{N}[(-1)^n C(nT)].
\end{aligned}
\label{O}
\end{equation}
DTC phase can be characterized by the long time-averaged value of $O$ with $O \sim 1$ for an ideal DTC phase. On the other hand, if the period-doubling oscillations in the correlation function are not persistent, the order parameter $O$ goes to zero. Panel [f] of Fig.~[\ref{diffJ_ct}] shows $O$ vs $h$ for various values of $J$ based on the autocorrelation function shown in other panels. For smaller values of $J$, system has weaker disorder and hence the Floquet eigenstates show non-ergodic MBL behaviour only up to small values of $h$. Therefore, the DTC phase for weakly interacting SK model, survives only up to small values of $h$. For example, for $J=0.5$, the order parameter becomes less than $0.25$ for $h \ge 0.2$ while for $J=5.0$, the order parameter remains of order $0.5$ even at a much larger value of the random transverse field.

Fig.~\ref{OP} shows the colorplot of the order parameter in the $h-J$ plane starting from a one domain wall initial state. Since there is no initial state dependence in the auto-correlation function, we expect the order parameter to result in the same phase diagram for any random initial state. For very weak interactions, DTC phase is seen only in a narrow regime close to $h=0$. As the strength of interaction $J$ increases, which also increases the strength of disorder in the system, DTC phase survives up to a much larger value of $h$ up to $h^\star$. The threshold value $h^\star$, below which the DTC phase is observed, is a monotonically increasing function of $J$ and is slightly more than the MBL transition point obtained from the analysis of Floquet spectra as shown in Section-II. For $h >h_c$, where the Floquet states do not show signature of MBL, but the LSR still remains non-ergodic, the DTC order parameter is still finite supporting the scenario that above $h_c$ system remains non-ergodic for a range of $h$. This non-ergodicity stablizes the DTC phase for $h_c < h < h^\star$. 

So far we have shown the results for the DTC order parameter for a fixed system size. It is also crucial to explore the system size dependence of the order parameter. As shown on middle panel of Fig.~\ref{OP}, for small values of $h$, $h <0.2$ for $J=2.0$, the DTC order parameter is almost system size independent which indicates the presence of a robust DTC phase in this regime. For larger values of $h$, the DTC order parameter shows a decay with increase in $L$. The rightmost panel shows $h^\star$ above which $O < 0.25$. The threshold field $h^\star$ is almost system size independent for weakly interacting SK model indicating the presence of a robust DTC phase in this parameter regime, while a slight reduction in $h^\star$ occurs with an increase in $L$ for larger values of $J$. 
\\
\\
{\bf{Robustness of the DTC Phase:}} To confirm whether the period doubling oscillations seen in autocorrelation function continue to much later times, we calculated time evolution of staggered magnetisation starting from the Neel state using exact diagonalization. Fig.~\ref{M_longt_h} shows the time evolution of $|M_z(t)|$ up to $10^9 T$ for three system sizes. For $h<h^{\star}$, after an initial rapid decay, $|M_{z}(t)|$ remains saturated for a very long time and this time increases exponentially with the system size. This diverging time scale for the saturation plateau in magnetization shows the rigidity of the DTC phase in periodically driven SK model. In complete contrast to this, for $h>h^\star$, magnetization does not show any plateau and monotonically decreases with time. Additionally, for $h>h^\star$, the decay time for $|M_{z}|$ remains the same as the system size increases.

\begin{figure*}[t]  
 \centering
    \subfloat{\includegraphics[width=0.69\columnwidth,height=0.56\columnwidth]{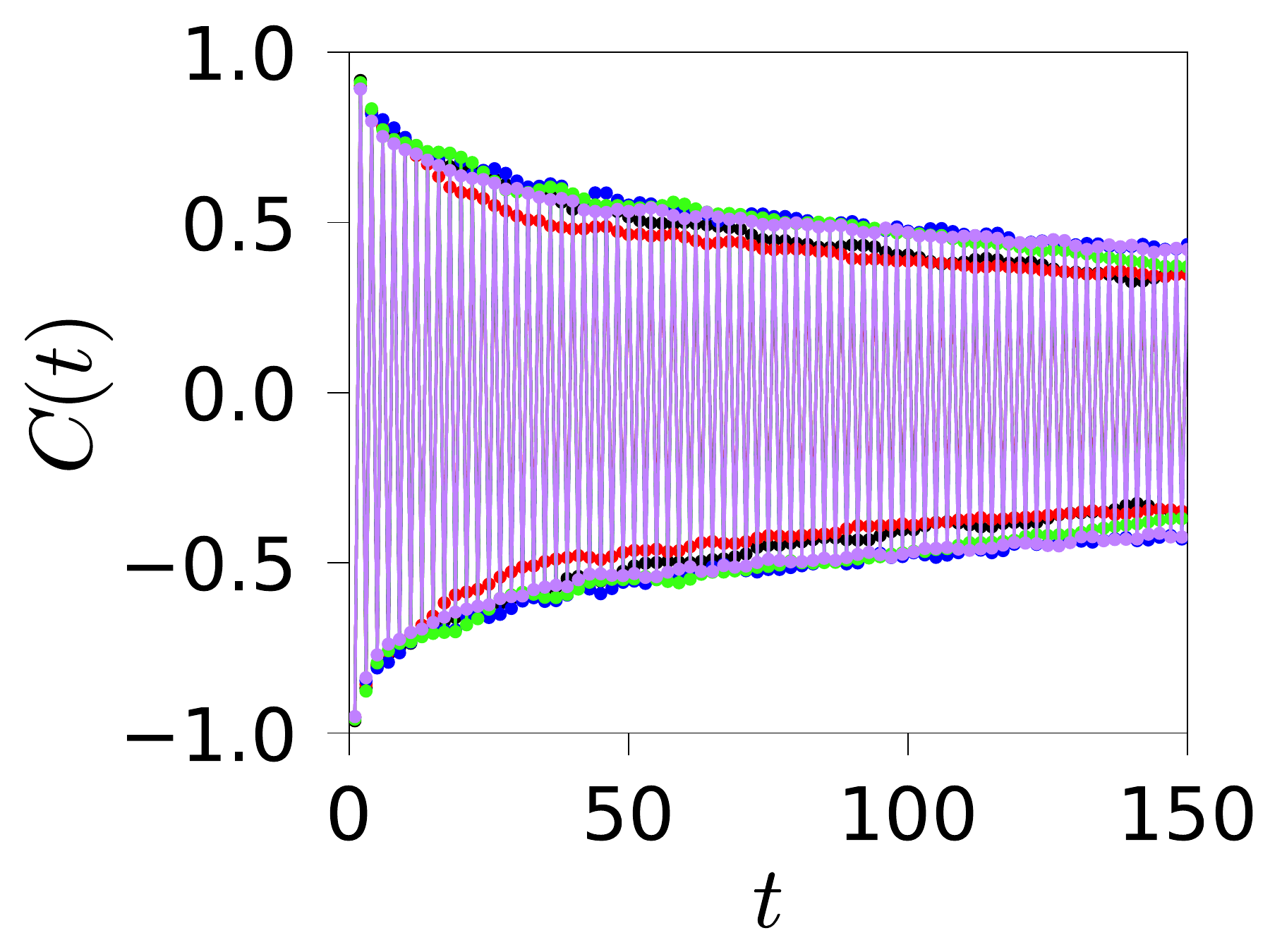}}
    \includegraphics[width=0.69\columnwidth,height=0.56\columnwidth]{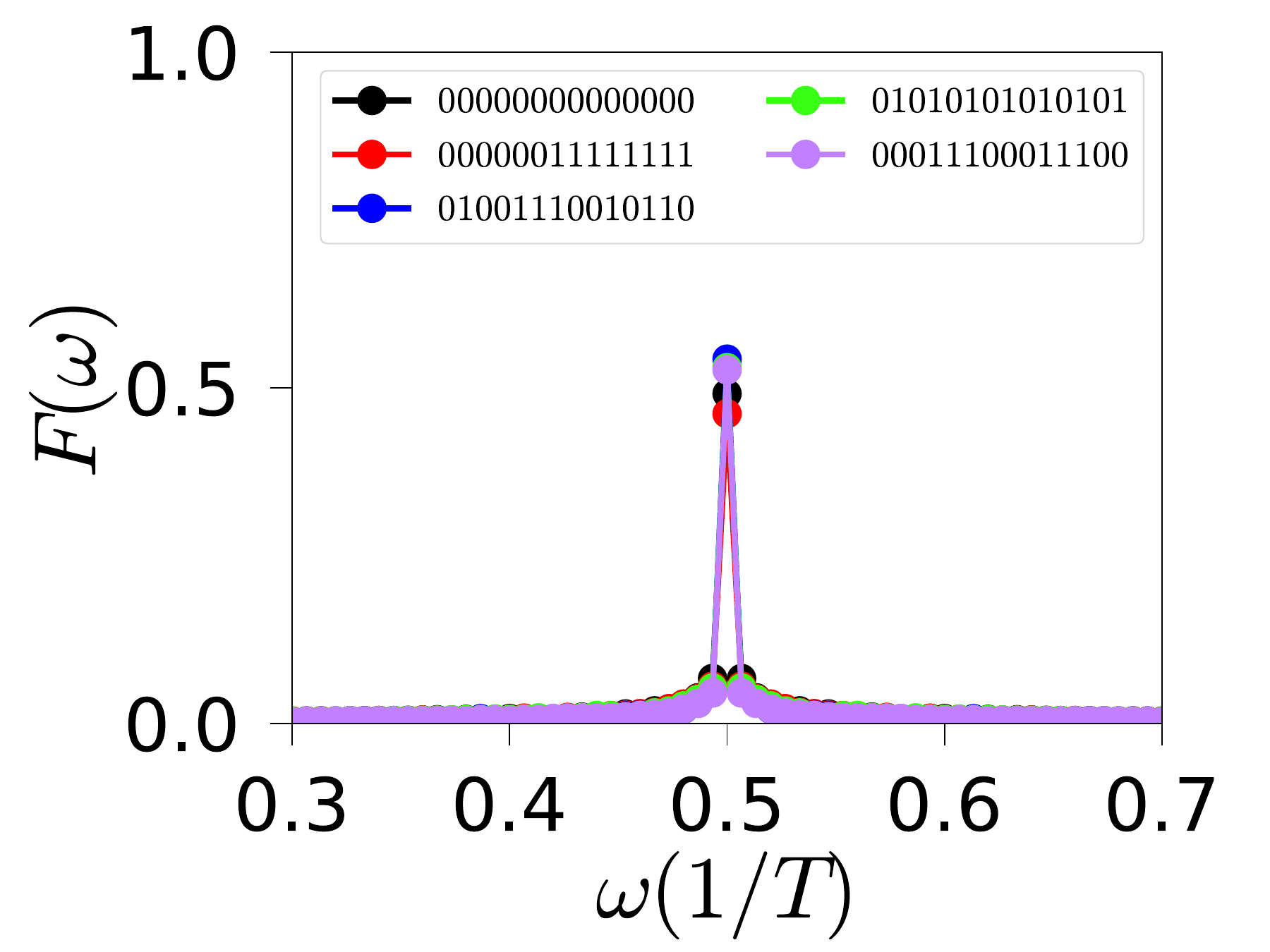}
    \includegraphics[width=0.65\columnwidth,height=0.57\columnwidth]{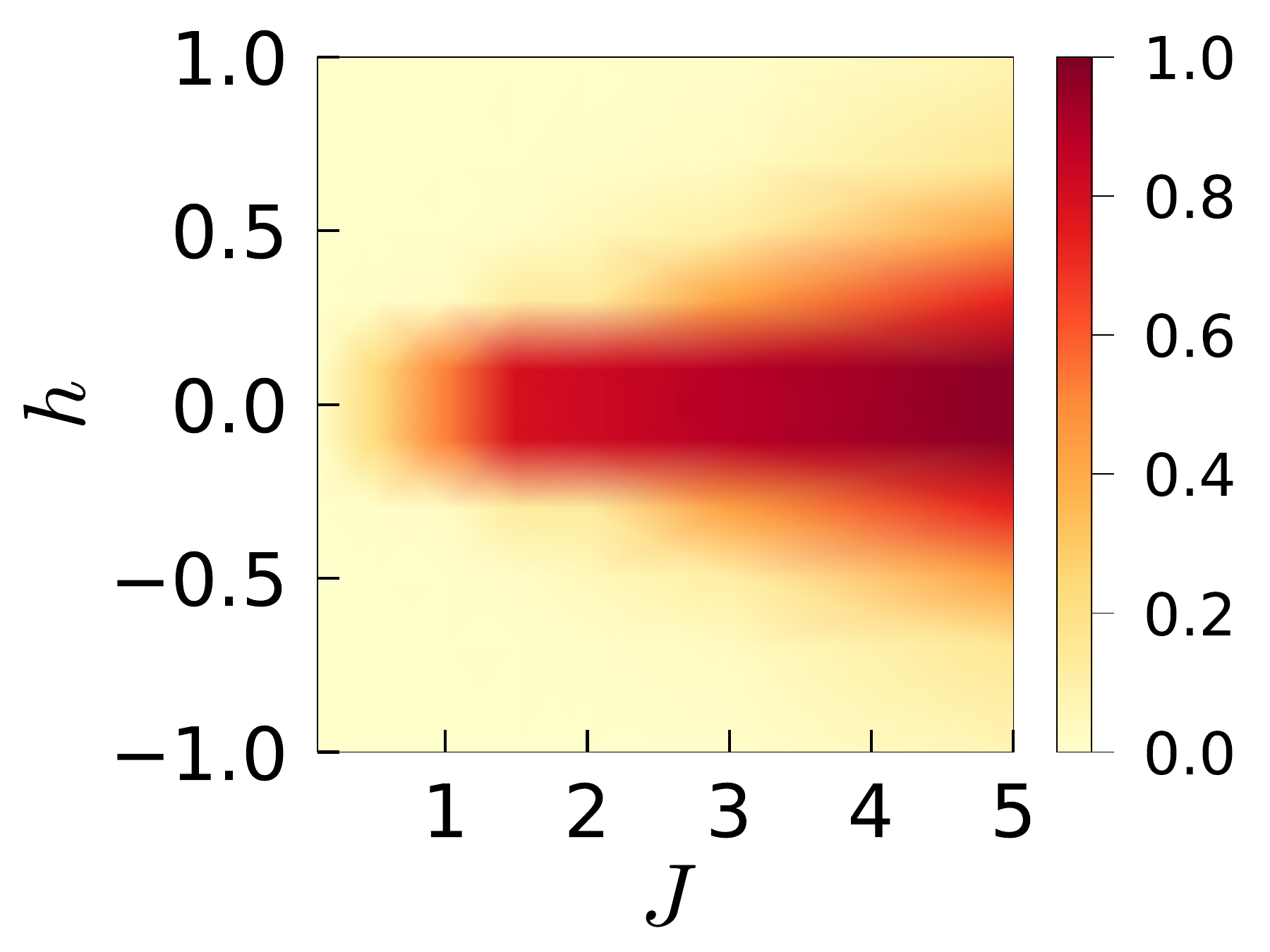} 
    \caption{The left panel shows auto-correlation function $C(t)$ for $J=2.0$ and uniform transverse field, $h=0.2$ for 5 random initial states. Middle panel shows the Fourier transform $F(\omega)$ of the correlations function which has a distinct peak at $\omega=0.5$. Rightmost panel shows the colour-plot of the order parameter $O$ obtained from initial state $|00000011111111\rangle$  in $J-h$ plane for the uniform $h$. The DTC phase survives for a narrow range of $h$ compared to the case of random transverse field. The data shown is for $L=14$ and have been averaged over 25 independent disorder configurations.}
    \label{unifh}
\end{figure*}

We would like to further emphasize that the regime in the $J-h$ plane where the correlation function shows persistent period doubling oscillations and the order parameter is large, is not only related the the non-ergodic phase of the Floquet eigenstates but also coincides to a great extent with the region where the eigenstates of the static Hamiltonian $H_2$ of the drive are in MBL phase. This is clearly seen from Fig.~\ref{se} in Appendix-A which shows Shannon entropy in $J-h$ plane. The regime where the DTC order parameter is more than half, coincides with the regime where the normalized Shannon entropy is less than half.  
This analysis shows that non-ergodicity caused by disorder in Ising spin couplings in the SK model stabilises the DTC phase despite long-range interactions among spins. 

{\bf System with Uniform Transverse Field:}
So far we have presented results for the random transverse field. We also studied the case of uniform transverse field $h^x_i=h$ to understand the role of inhomogeneity in this local field. As shown in Fig.~\ref{unifh} correlation function shows period-doubling oscillations for all the initial states studied for the uniform transverse field as well, but here the DTC phase exists for a narrow range of $h$ around $h=0$ (Fig.~\ref{unifh}) compared to the case of random transverse field. 
Based on this analysis, we conclude that although SK model has fully random Ising interactions, having randomness in the local transverse field is crucial for stabilizing a broad DTC phase.  This is an important conclusion because generally it is believed that disorder in Ising-even terms rather than in local longitudinal or transverse field stabilizes a DTC phase~\cite{Ippoliti2021}.
\begin{figure*}[]
    \centering
    \includegraphics[width=2.0\columnwidth]{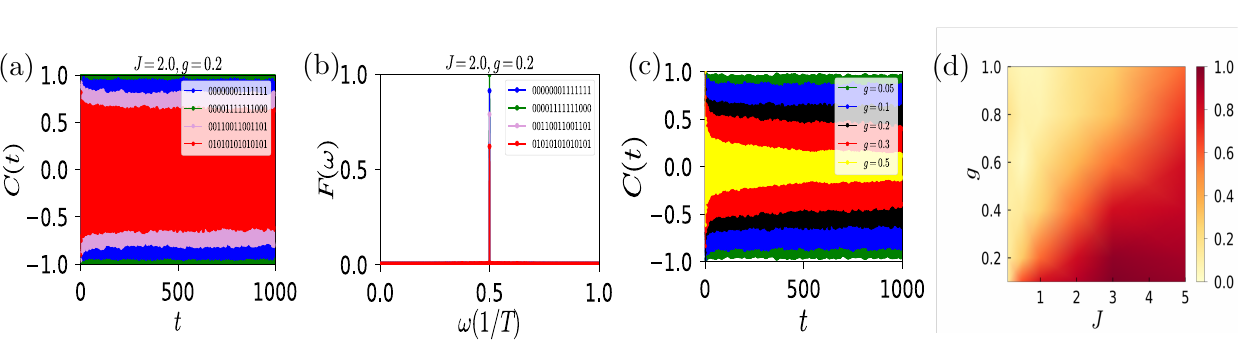}

    \caption{Panel[a]: Auto-correlation function $C_{L/2}(t)$ for $J=2.0$ and random $XY$ coupling with strength $g=0.2$ for various initial states. [b]The corresponding Fourier transform $F(\omega)$ has a pronounced peak at $\omega=0.5$.[c] Auto-correlation function for the Neel state being the initial state for varying strength of random $XY$ coupling, $g$, for a fixed $J=2.0$. The data shown are for $L=14$ and averaged over 100 disorder configurations[d] Order parameter for the Neel state in $J-g$ plane obtained by averaging over time interval $0-300$.  Data shown are for $L=14$ and has been averaged over 25 disorder configurations.}
    \label{rang}
\end{figure*}

\begin{figure}[]
    \centering
    \subfloat{\includegraphics[width=0.52\columnwidth,height=0.4\columnwidth]{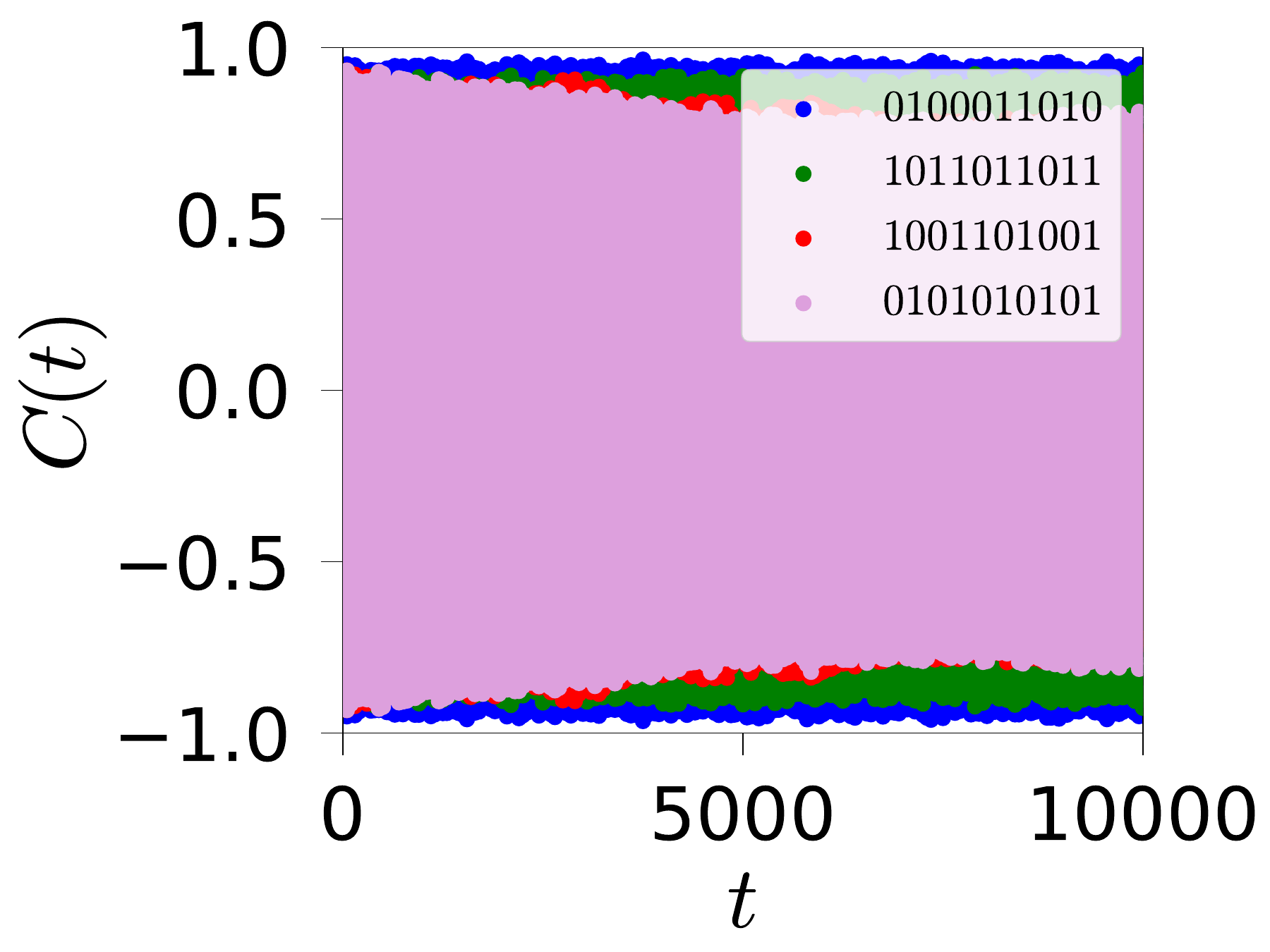}}
    \includegraphics[width=0.52\columnwidth,height=0.4\columnwidth]{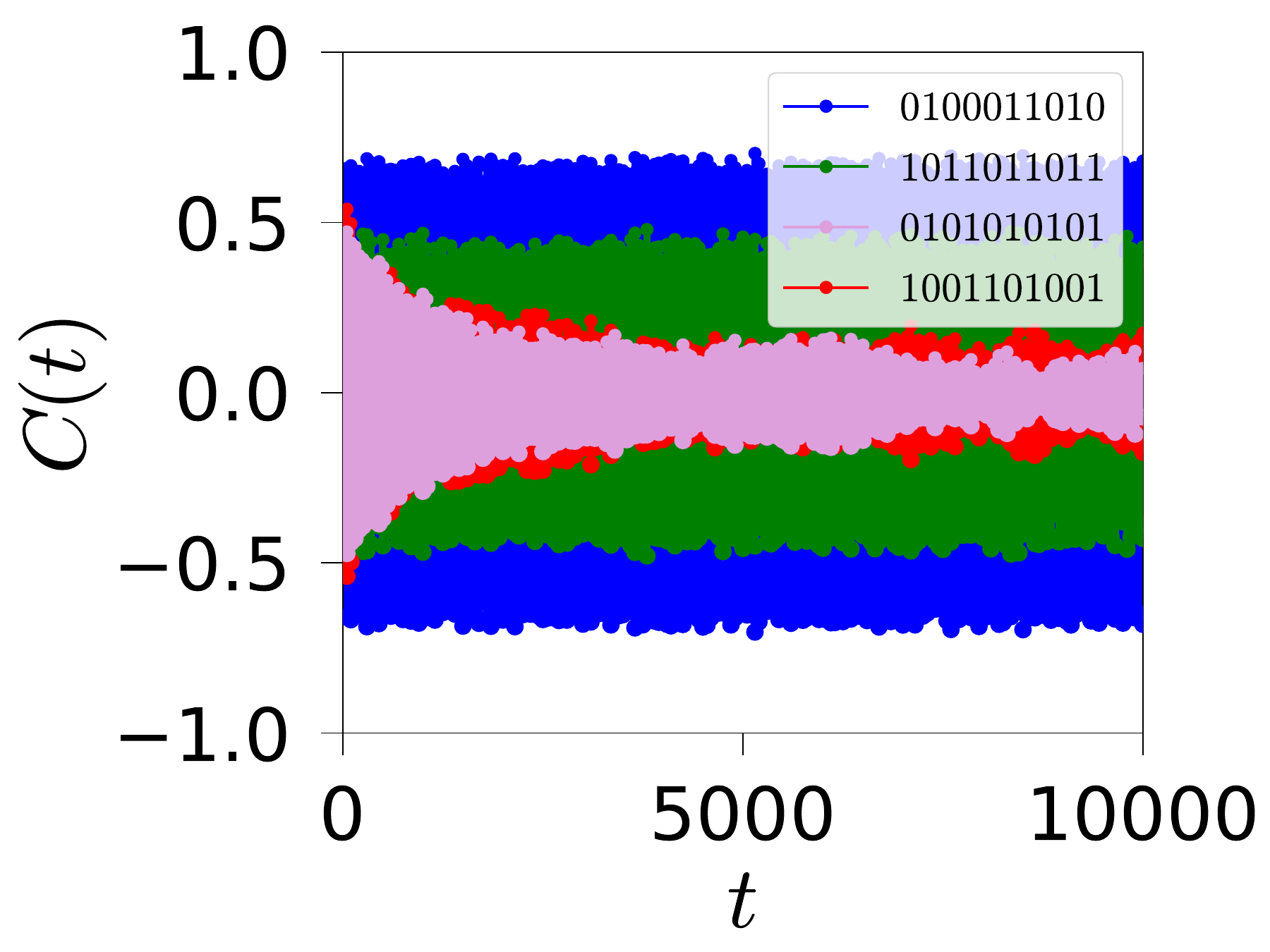}
    \caption{Correlation function derived for 10000 periods - SK interaction with $J=2.0$, L=10, 4 different initial states, avg over 100 disorder configs with random xxyy interaction with (left) $g=0.1$ (right) $g=0.4$}
    \label{longtime_g}
\end{figure}

\subsection{Driving protocol-II}
In this Section we explore the DTC phase in the Floquet Hamiltonian in Eqn.~\ref{d2} with random $XY$ coupling on the nearest-neighbour bond to induce quantum dynamics in the SK model. The question of interest is whether the broad regime of non-ergodic phase (discussed in Section II) will help in stabilizing a broader DTC phase in this case or the faster dynamics induced by the $XY$ term will destablize the DTC phase.

We first explore the system with random $g_i$ drawn from a uniform distribution $g_i \in[0,g]$. We study the spin-spin autocorrelation function $C(t)$ starting from a random initial state. 
First panel in Fig.~\ref{rang} shows the auto-correlation function at the $L/2^{th}$ site of the chain for $J=2.0$ and $g=0.2$ for various initial states.  All initial states show persistent period doubling oscillations which is also reflected in the Fourier transform $F(\omega)$  of the correlation function which shows a peak at $\omega=0.5(2\pi/T)$ shown in the second panel of Fig.~\ref{rang}.  As the strength of $g$ increases, $C(t)$ decays faster at short time scales due to faster dynamics. This suppresses the correlation function making it less than one in the short time regime as shown in the third panel of Fig.~\ref{rang}, however $C(t)$ retains this amplitude for a very long time without any further decay of period doubling oscillations. Therefore, for a given strength of SK coupling $J$, as the strength of random $XY$ coupling is increased, the DTC phase survives for a broad range of $g$ and eventually the DTC order melts for $g >g^\star$ as shown in the rightmost panel of Fig.~\ref{rang} which shows colorplot of the order parameter $O$ for the Neel state. 

Comparing the dynamics in this model with that in driving protocol-I, it seems that having a $XY$ coupling provides a DTC phase over a broader range of parameters than having a transverse field in $H_2$. We believe that the wider DTC phase in $J-g$ plane is due to the enhanced non-ergodicity of the eigenstates of the Floquet system under consideration. As an illustration, we have shown the results for $J=2.0$ starting with the Neel state in the third panel of Fig.~\ref{rang} for various values of $g$. Even for $g=0.5$ the system shows persistent period-doubling oscillations while in the case of random transverse field, $C(t)$ decays significantly with time for $h \ge 0.5$. 
But there are initial state dependence in the autocorrelation function for the model under protocol-II, which we will discuss below. 

\begin{figure*}[t]  
 \centering
    \subfloat{\includegraphics[width=1.7\columnwidth]{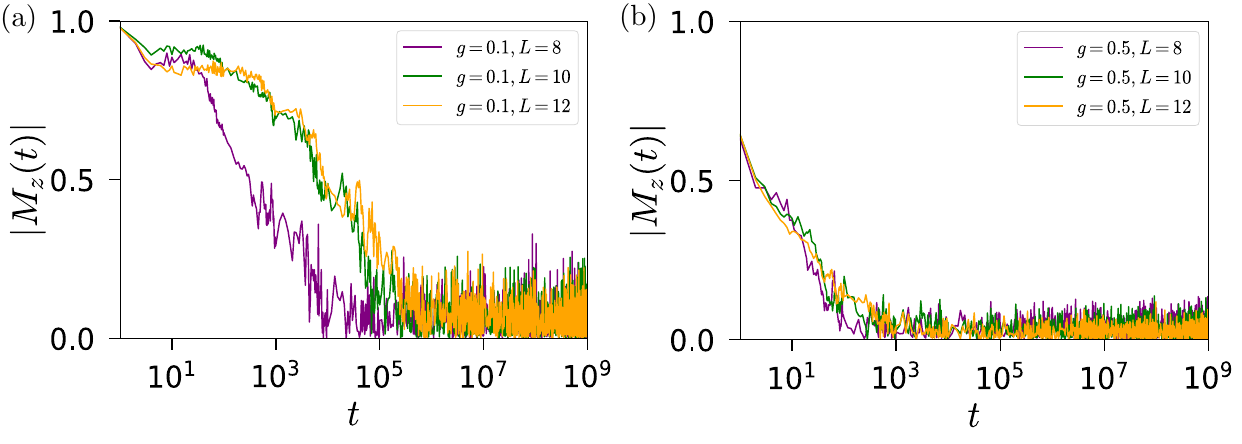}}
    \caption{Staggered magnetization ($|M_z(t)|$) for the Neel state as initial state for the driving protocol-II  with random spin flip term for three systems sizes. The left panel shows the results for $g=0.1$ and the right panel shows the results for $g=0.5$. The data shown are for $J=2.0$ and has been averaged over 50 independent disorder configurations.}
    \label{M_longt_g}
\end{figure*}

{\bf Robustness of the DTC Phase}: To check the robustness of the above observation, we calculated the correlation function $C(t)$ up to larger time scales of the order of $10^4$. As shown in Fig.~\ref{longtime_g}, for small values of $g$ the autocorrelation function shows period doubling oscillations for very large time scales up to $10^4$ for all the initial states however a clear initial state dependence is seen in large time response for systems with larger values of $g$ as shown in the right panel of Fig.~\ref{longtime_g} in the sense that amplitude of $C(t)$ in the long time limit depends on the initial state. Interestingly enough, even then for all the initial states, period doubling oscillations still persist though with a much smaller amplitude of the correlation function for some of the initial states. Thus, although for some of the initial states, the DTC order is robust for a long time for a wide range of $g$, for some other initial states the DTC order decays fast as $g$ increases.

We further calculated time evolution of staggered magnetisation starting from the Neel state. Fig.~\ref{M_longt_h} shows time evolution of $|M_z(t)|$ up to $10^9 T$ for three system sizes.
There is a clear contrast from the case of random transverse field in the sense that $|M_z(t)|$ does not show broad plateau even for small values of $g$. But for small values of $g$, after the initial rapid decay, magnetization remains almost flat for $t  \le 5*10^3$ for larger sizes (e.g. for $L=12$). The width of the plateau increases with the system size but not as much as in the case of the system with random transverse field. 
For larger values of $g$, for all the system sizes studied, magnetization decays to zero for $t \sim 10^3$. Note that the Neel state shows the fastest decay of the autocorrleation function as $g$ increases as shown in Fig.\ref{longtime_g}. For some of the other initial states, $|M_z(t)|$ remains robust for much larger time scales which are comparable to the timescales in the system with random transverse field. 

We further analyze the case where $g_i=g$ is uniform at all sites details of which are provided in Appendix B. For uniform $g$ the autocorrelation function $C(t)$ decays even faster at short time scales compared to the random spin-flip case. For a fixed strength of the SK coupling as the strength of $g$ increases, the correlation function decays for smaller values of $g$ compared to the case of a random $XY$ term resulting in a narrower DTC phase in the $J-g$ plane. This analysis demonstrates that randomness in the $XY$ couplings supports the period doubling oscillations in the spin-correlation function. 

\subsection{Driving protocol-III}
In the third driving protocol we study the following system 
\begin{equation}
H_f(t) = 
\begin{cases} 
H_1 = (\frac{\pi}{2}-\epsilon)\sum\limits_i \sigma_i^x, & 0 < t < 1 \\ 
H_2 = \sum\limits_{i<j}J_{ij}\sigma_i^z\sigma_j^z, & 1 < t < 2
\end{cases}
\label{d3}
\end{equation}

In this case, it is possible to obtain an analytical expression for the time evolution operator for each step of the drive as follows,
\begin{equation}
    U_1=e^{-iH_1}=e^{-i(\frac{\pi}{2}-\epsilon)\sum\limits_{i=1}^L\sigma_i^x}=\prod\limits_{i=1}^L e^{-i(\frac{\pi}{2}-\epsilon)\sigma_i^x}
\end{equation}
Taking the Taylor expansion of exponential and making use of properties of Pauli matrices, the equation reduces to
\begin{equation}
    U_1=\prod\limits_{i=1}^L [cos(\frac{\pi}{2}-\epsilon)\mathbb{I}-isin(\frac{\pi}{2}-\epsilon)\sigma_i^x]
\end{equation}
For the second part of the Hamiltonian, we have,
\begin{equation}
    U_2 = \prod\limits_{1\leq i<j \leq L}[cos(J_{ij})\mathbb{I}-isin(J_{ij})\sigma_i^z\sigma_j^z]
\end{equation}
\begin{figure*}[]
    \subfloat{\includegraphics[width=0.55\columnwidth,height=0.45\columnwidth]{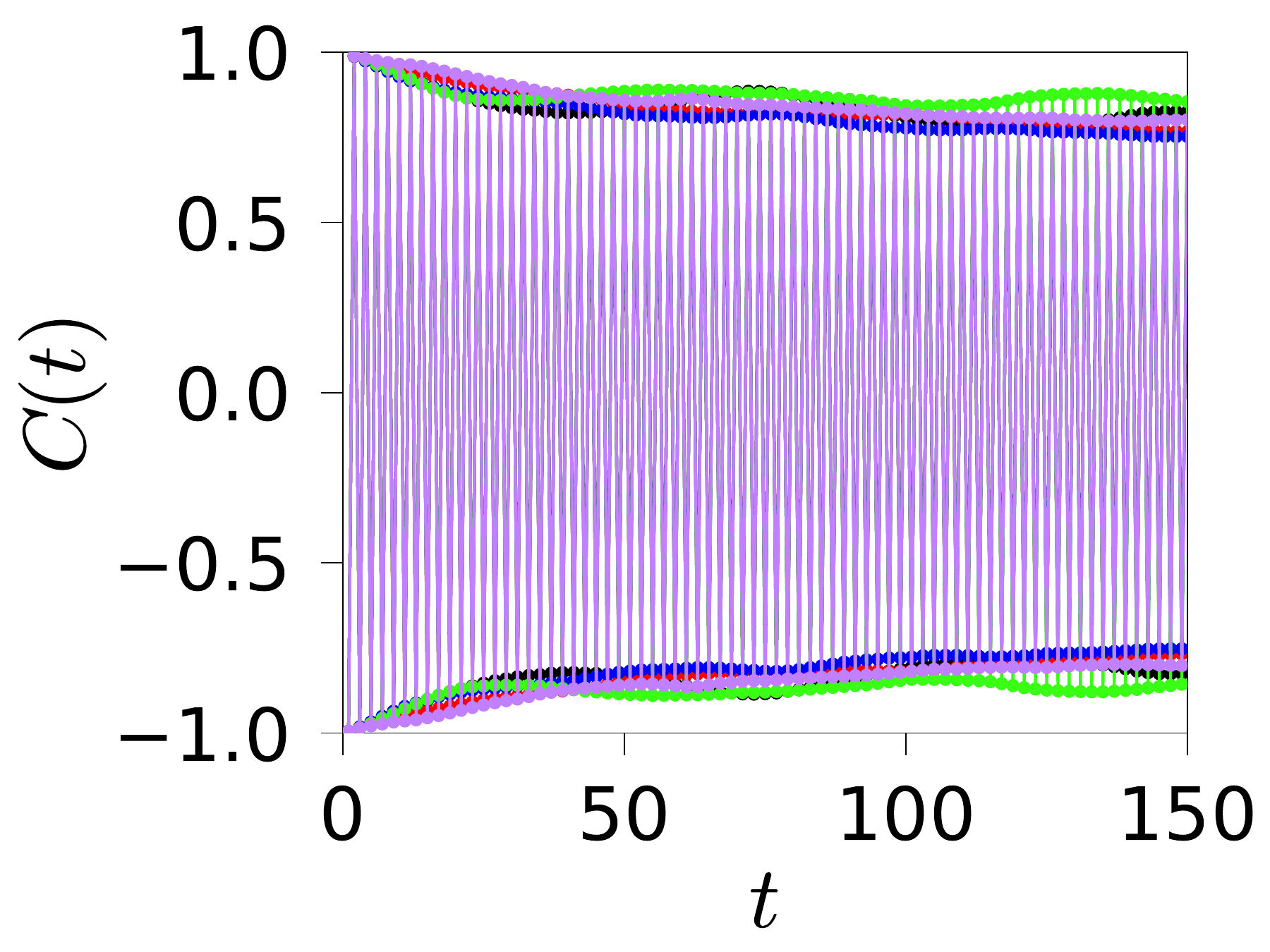}}
    \includegraphics[width=0.62\columnwidth,height=0.45\columnwidth]{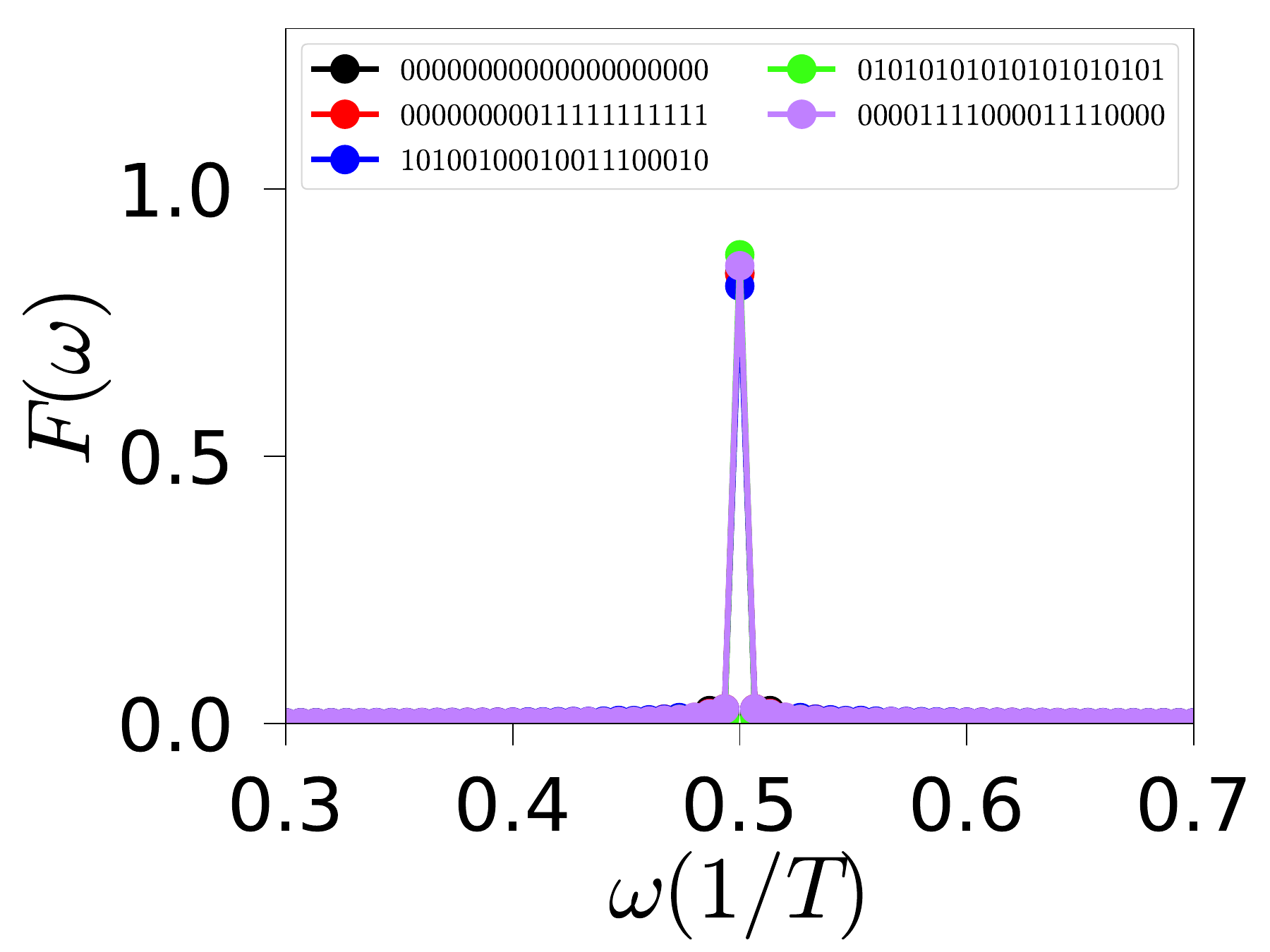}
        \includegraphics[width=0.525\columnwidth,height=0.45\columnwidth]{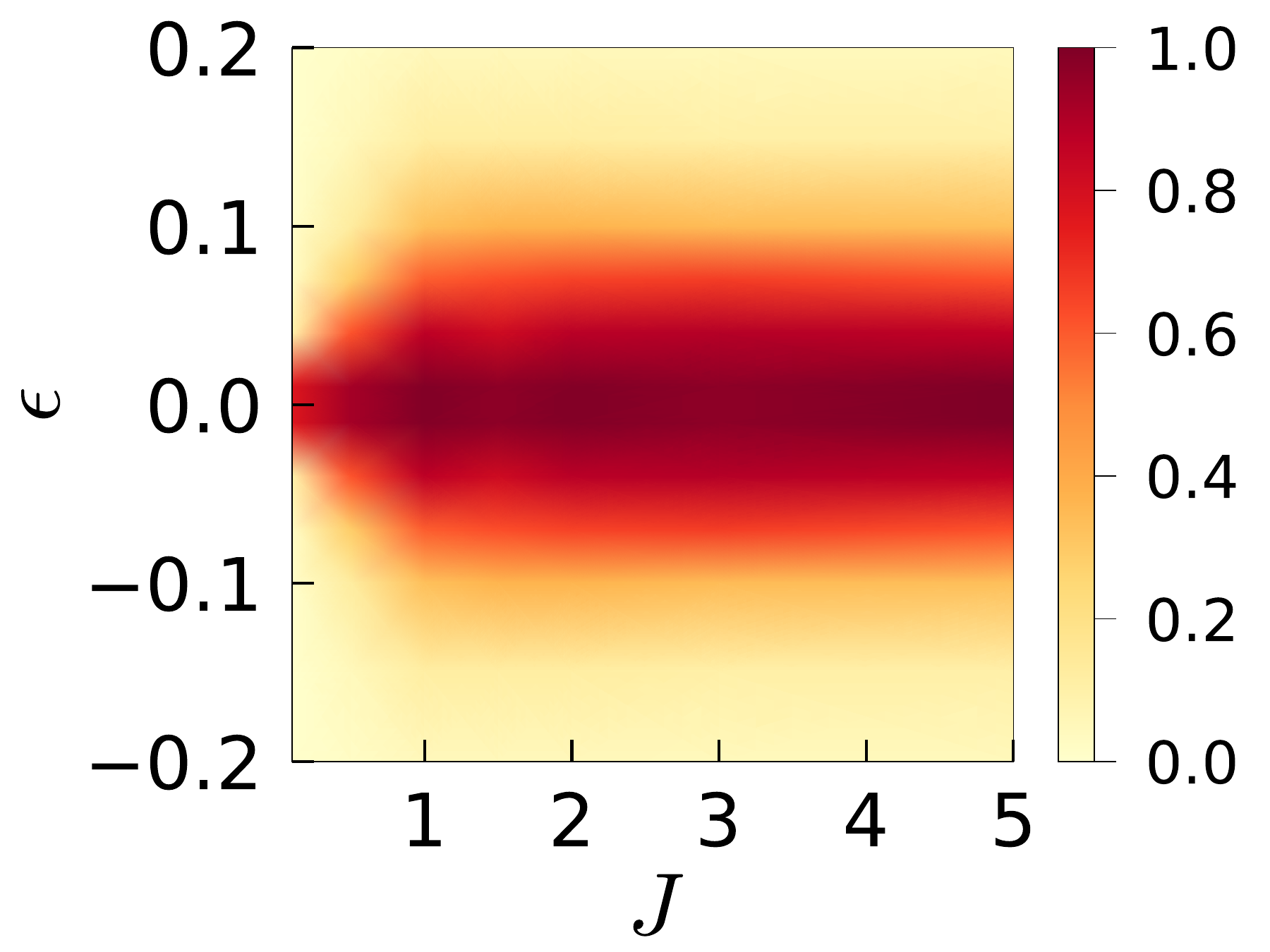}
    \caption{Auto-correlation function $C(t)$ together with its Fourier transform for spin chain with $L=20$, $J=2.0$ and $\epsilon=0.05$ for the driving protocol-III.  The rightmost panel shows the phase diagram generated from the time-averaged (300 periods) order parameter $O$ starting with initial state $|00000000011111111111\rangle $. Data shown has been averaged over 50 independent disorder configurations.}
    \label{eps}
\end{figure*}

As shown in Fig.~\ref{eps}, there is hardly any initial state dependence of the spin-spin autocorrelation function. Even for finite non-zero $\epsilon$, for any random initial state correlation functions shows period doubling oscillaltions for a very long time indicating the presence of a DTC  phase. The width of the DTC phase does not change much with $J$ here in contrast to the driving protocols -I and II. For any non-zero value of $J$, the width is more or less the same along the $\epsilon$ axis being around $-0.1 < \epsilon < 0.1$ as shown in Fig.~\ref{eps}.

We have also compared the unitary time evolution dynamics of the SK model with other spin models of long-range interactions. We demonstrated that systems with uniform power-law interactions can host only prethemal DTC phase with strong initial state dependence even in the presence of random local fields. Details of this are given in Appendix C.

\section{Conclusions}
\label{sec-5}

Time crystals are fascinating phases of matter that emerge in out-of-equilibrium quantum many-body systems. They have potential uses in quantum computing, quantum sensing, and quantum storage. It is natural to wonder what kind of systems can support a discrete-time crystal phase. According to general understanding, a non-ergodic short-range interacting system with MBL can support the formation of a discrete-time crystal (DTC) phase when periodically driven under certain conditions, though prethermal DTC can occur even in clean and ergodic systems.   
In this work, we explored the question of whether there are long-range interacting spin models capable of hosting a stable robust DTC phase and presented a positive answer. We showed that a quantum variant of the Sherrington-Kirkpatrick model of Ising glass, with all-to-all spin interactions and totally random couplings, provides an excellent framework for realizing a stable DTC under diverse driving protocols.
We demonstrated that the width of the DTC phase is directly related to the width of non-ergodic phase of Floquet eigenstates of the system. 
The DTC phase is seen in the non-ergodic regime of the Floquet Hamiltonian, but not the entire non-ergodic regime shows MBL.

We considered unitary time evolution dynamics of a periodically driven SK model under various driving protocols and found that having a random transverse field to induce quantum dynamics in the SK model provides a robust DTC phase in which the autocorrelation function shows persistent period doubling oscillations for a long time and has no dependence on the initial state. In contrast to this, for a uniform transverse field $h$, the width of the DTC phase is much narrower in the $J-h$ plane. Interestingly, having a random spin-flip $XY$ term to induce quantum dynamics in the SK model is more beneficial in providing a broad  DTC phase. But the amplitude of these oscillations shows initial state dependence for larger strengths of the random $XY$ term. 


We believe that our work provides a new pathway for realizing stable robust-time crystals, and it would be interesting to see some of these theoretical proposals become realized in experiments.  

\section{Acknowledgments} A.G. acknowledges support from
the Anusandhan National Research Foundation, Government
of India (Grant No. ANRF/ARGM/2025/000622/TS). We would like to acknowledge the National Supercomputing Mission (NSM) for providing computing resources of ‘PARAM RUDRA’ at S.N. Bose National Centre for Basic Sciences, which is implemented by C-DAC and supported by the Ministry of Electronics and Information Technology (MeitY) and Department of Science and Technology (DST), Government of India.
Aarya Bothra would also like to acknowledge the support provided by
the High Performance Computing (HPC) Cluster (Kepler) facility, maintained by the Department of Physical
Sciences, IISER Kolkata.

\appendix 
\section{Appendix A: Shannon Entropy for SK model with random transverse field}
\begin{figure}[h]
    \subfloat{\includegraphics[width=0.45\columnwidth]{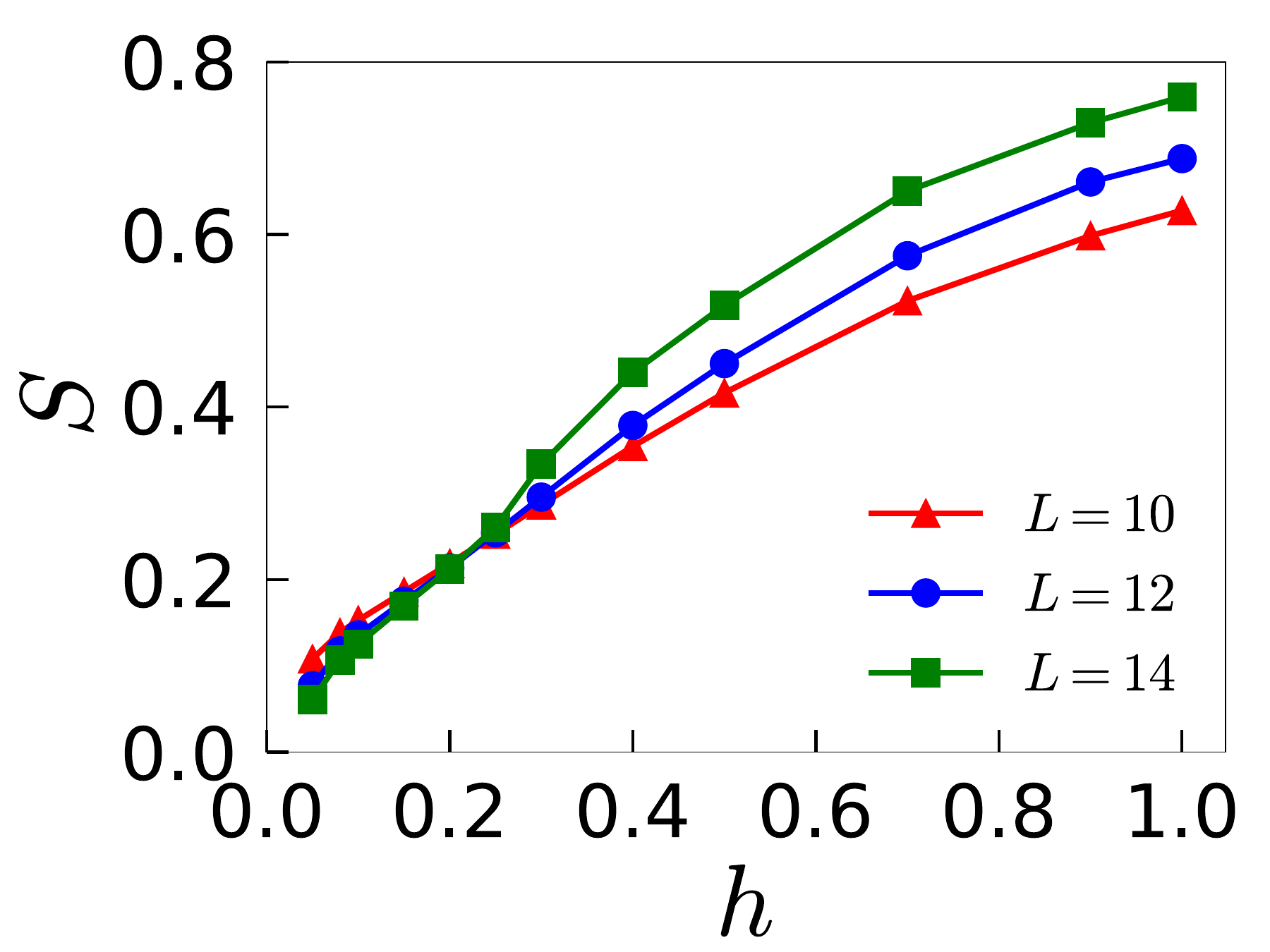}}
      \includegraphics[width=0.45\columnwidth]{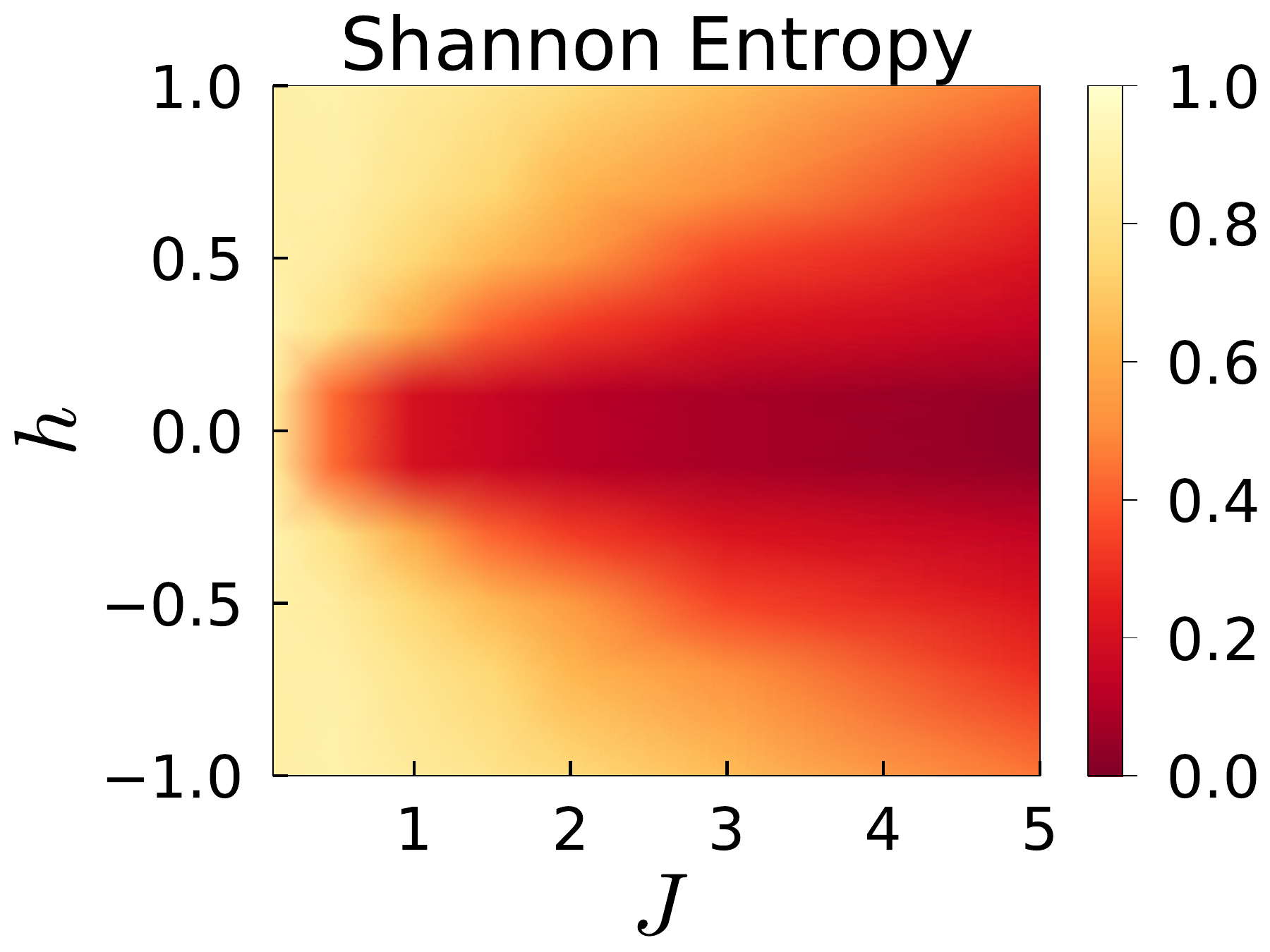}
    \caption{Shannon entropy obtained from eigenstates of $H_2$ in the driving protocol-1 for the SK model with random transverse field $h_i^x \in [0,h]$. Data shown are averaged over 2000 (L=10), 1000 (L=12), 100 (L=14) disorder configurations respectively. Left plot corresponds to the Shannon entropy for $J=2.0$ for three system sizes. Right plot is the colour plot for the shannon entropy in $J-h$ plane for $L=14$ and has been averaged over 50 disorder configurations.}
      \label{se}
      \end{figure}

In the main text we have done MBL analysis for the Floquet Hamiltonian. Here we provide details of the Shannon entropy obtained from the eigenstates of $H_2$ Hamiltonian for the case of SK model with random transverse field.  
 As shown in the Fig.~\ref{se}, as the strength of the random transverse field increases for a fixed strength of SK interactions, the normalized Shannon entropy increases from zero to one showing the delocalization induced by the transverse field. The regime in the $J-h$ plane where the correlation function shows persistent period doubling oscillations and the order parameter is more than half, coincides to a great extent with the region where the normalized Shannon entropy is less than half. This can be understood as follows. If eigenstates of $H_2$ are localized with small values of Shannon entropy, then an initial state will have overlap with only a small fraction of eigenstates and such a state will not spread much under unitary time evolution with $H_2$ resulting in larger spin auto-correlation function and the DTC order parameter within the given driving protocol. On the contrary, if eigenstates are extended having large number of basis states contributing to them, initial state will also have overlap with a large fraction of eigenstates. Such a state will show a larger spread in the eigenbasis as it evolves in time resulting in decay of correlation function. This indicates crucial role of highly localized eigenstates of $H_2$ in stabilizing the DTC phase.

\section{Appendix B: Autocorrelation function for uniform spin-flip in driving protocol-II}
In the main text we have presented unitary time evolution dynamics for periodically driven SK model under driving protocol-II for the case of random spin-flip terms. Here, we study the autocorrelation function for the case of uniform spin-flip term. 
\begin{figure}[h]
    \subfloat{\includegraphics[width=1.0\columnwidth]{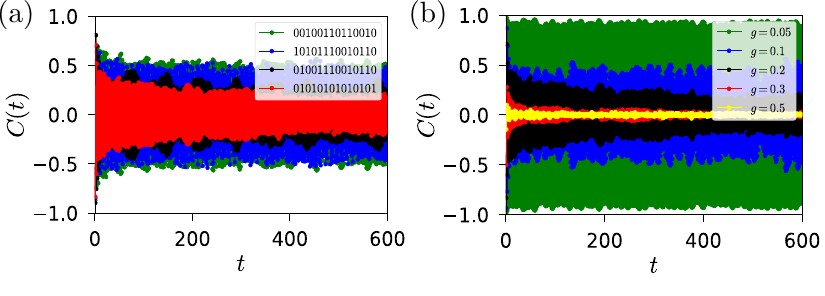}}
    \caption{Auto-correlation function $C_{L/2}(t)$ for the periodically driven SK model under driving protocol-II with uniform $XY$ coupling $g =0.2$. Right panel shows the effect of varying spin-flip term for the initial state being the Neel state. Data is shown for a spin chain with $L=14$, $J=2.0$ and has been averaged over 10 independent disorder configurations.}
    \label{unifg}
\end{figure}

First panel of Fig.~\ref{unifg} shows $C(t)$ vs $t$ for $J=2.0$ and $g=0.2$ for various initial states. For all the initial state there is a fast decay of $C(t)$ at very short time but after that the correlation function remains almost of the same amplitude.  There is a dependence on the initial states and for all the cases $C(t)$ for uniform $XY$ term is smaller than that for the random $XY$ term. But for all the cases, period doubling oscillations survive for a long time which is also reflected in a prominent peak in the Fourier transform $F(\omega)$ of the correlation function. The rightmost plot in Fig.~\ref{unifg} shows the correlation function for various values of $g$ for $J=2.0$ starting from a Neel state. As $g$ increases, $C(t)$ decreases becoming almost zero for $g=0.3$ resulting in a much narrower DTC phase compared to the system with a random $XY$ term. 

\section{Appendix C: Comparison of SK model with long-range interacting systems having uniform interactions}
In this Section we compare the unitary time evolution dynamics of periodically driven SK model with other spin models of long-range interactions. To be specific, we consider the case of Ising coupling with the power-law form $\sum\limits_{i<j}\frac{J}{|i-j|^\alpha}\sigma_i^z\sigma_j^z$ with a uniform coupling $J$. For $\alpha \le d$ where $d$ is the physical dimensionality of the system, the interactions are long-range in nature while for $\alpha > 2d$ interactions are considered to be of short-range. We study the case for $\alpha=1$ and $\alpha=2$ here for one-dimensional system.  Clean systems with power-law interactions are known to show prethermal DTC~\cite{Machado2020,Pizzi2021_higher}. But here we are interested in the dynamics of power-law interacting spin systems where the system has disorder either through the random transverse field or through the random $XY$ term. 

In driving protocol-I, replacing the SK interaction with uniform power-law interactions, gives the following Floquet Hamiltonian $H_f(t)$:
\begin{equation}
H_f(t) = 
\begin{cases} 
H_1 = \frac{\pi}{2}\sum\limits_i \sigma_i^x, & 0 < t < 1 \\ 
H_2 = \sum\limits_{i<j}\frac{J}{|i-j|^\alpha}\sigma_i^z\sigma_j^z + \sum\limits_ih_i^x\sigma_i^x, & 1 < t < 2
\end{cases}
\label{d1-LR}
\end{equation}

\begin{figure*}[]
    \vskip0.4cm
    \subfloat{\includegraphics[width=1.7\columnwidth]{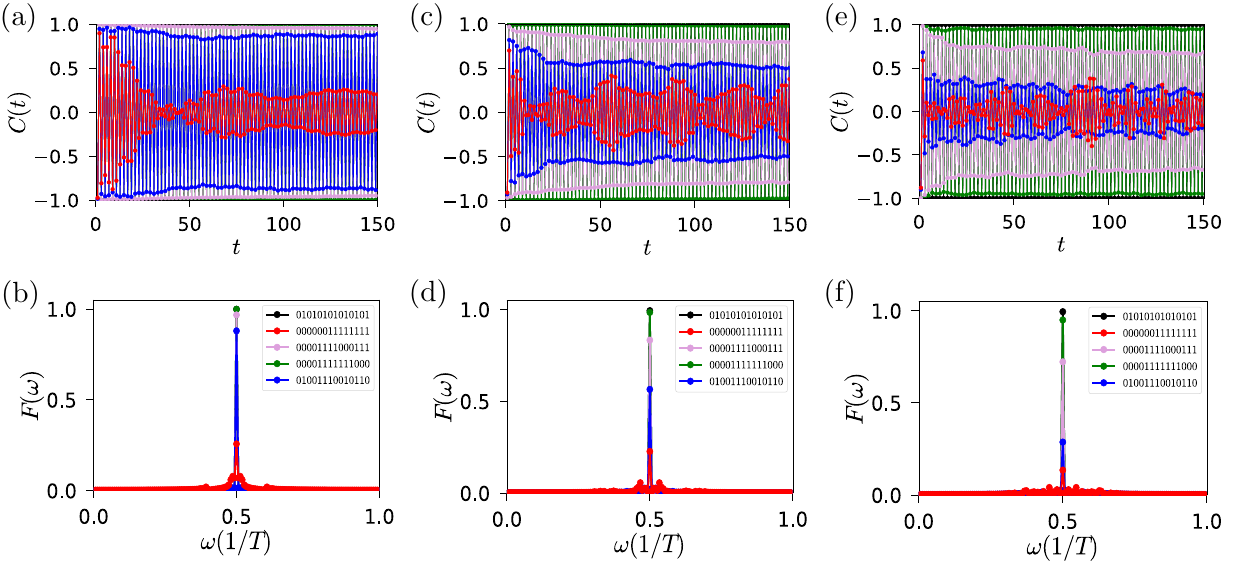}}

    \caption{Auto-correlation function $C(t)$ together with its Fourier transform $F(\omega)$ for power-law interacting system under the driving protocol-I at $J=2.0$ and random field $h_i \in [0,h]$. Data shown for (a-b) $\alpha=1.0$ and $h=0.2$, (c-d) $\alpha=1.0$ and $h=0.4$, (e-f) $\alpha=2.0$ and $h=0.4$. Random initial states were chosen for comparision for a spin chain of size $L=14$ with averaging done over 25 disorder configurations for each.}
    \label{LR1}
\end{figure*}

\begin{figure}
 \hspace{-1.1cm}
\subfloat{\includegraphics[width=1.1\columnwidth]{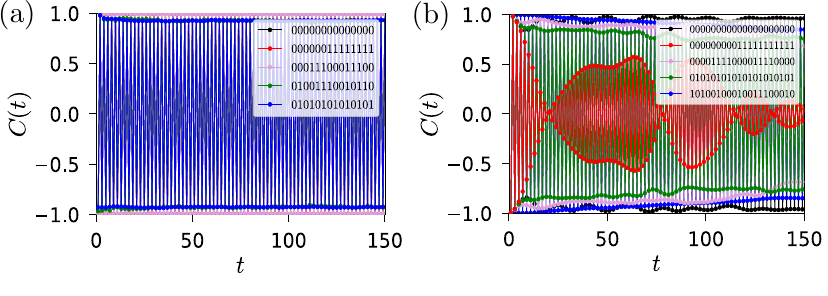}}
   \caption{
   Autocorrelation function $C(t)$ for system with uniform power-law interactions with $\alpha=1$. The left panel shows the result for $\alpha=1$ for the system with random spin flip terms of strength $g=0.2$ while the right panel is for the third driving protocol with $\epsilon=0.05$ and $\alpha=2.0$. The data shown are for $J=2.0$ and $L=14$. }
   \label{power_2}
\end{figure}

Starting from a random initial state, for unitary time evolution under this model, the autocorrelation function, $C(t)$, shows a strong dependence on initial states with a large number of initial states showing decay in oscillations even at short time. In Fig.~\ref{LR1} we have shown $C(t)$ for the drive with random transverse field $h_i^x \in[0,h]$ for $J=2.0$. For both $\alpha=1$ and $\alpha=2$, $C(t)$ lacks persistent period doubling oscillations, which is also reflected in many small peaks around $\omega=0.5W$ with $W = 2\pi/T$ in the Fourier transform of $C(t)$. This is in complete contrast to the case of SK model in the presence of random transverse field where there is no initial state dependence making the DTC phase very robust. 
We also explored the third driving protocol with an offset $\epsilon$ in the $\pi$ pulse for systems  with power-law interactions and saw a decay in auto-correlation function for some initial states as shown in Fig.\ref{power_2} and shows consistency with the known literature.
We finally analysed the driving protocol-II for power-law interacting spin model. 
In the case of random spin-flip terms, persistent period doubling oscillations are seen for $\alpha=1$ as well as $\alpha=2$ for very long time. Though these will suffer from the issue of weak initial state dependence as in the case of SK model with random spin-flip terms for large values of $g$, still it is possible to have persistent period doubling oscillations for almost all initial states in this case. 
Therefore, in systems with uniform power-law interactions, there is always an initial state dependence and only a prethermal DTC phase can be realized in the driving protocol-1 and III when the system has disorder through local terms like local random transverse field . Only in the protocol-II with random spin-flip terms it is possible to have period doubling oscillations for almost all initial states for a long time though the amplitude of oscillations are initial state dependent for large $g$.   

 


\bibliography{references.bib}

\end{document}